\documentclass[twoside,lettersize,journal]{IEEEtran}
\usepackage{amsmath,amsfonts}
\usepackage{algorithmic}
\usepackage{algorithm}
\usepackage{array}
\usepackage[caption=false,font=normalsize,labelfont=sf,textfont=sf]{subfig}
\usepackage{threeparttable}
\usepackage{textcomp}
\usepackage{stfloats}
\usepackage{url}
\usepackage{soul}
\soulregister{\cite}7
\usepackage{soul}
\usepackage{hyperref}
\usepackage{refcount}
\usepackage{verbatim}
\usepackage{graphicx}
\usepackage{cite}
\usepackage{booktabs} 
\usepackage[table,xcdraw]{xcolor}
\usepackage{diagbox}
\usepackage{multirow} 
\usepackage{lipsum} % 用于生成示例文本
\usepackage{booktabs} % 用于漂亮的表格线
\usepackage[utf8]{inputenc}
\usepackage{fontawesome5} % 导入 fontawesome 宏包
\usepackage[many]{tcolorbox} % for summary box
\usepackage{orcidlink}
\newtcolorbox{boxK}{
    top=2pt,
    bottom=2pt,
    left=2pt,
    right=2pt,
    boxrule = 0pt,
    toprule = 0pt, % top rule weight
}

\begin{document}

\title{Automated Commit Message Generation with Large Language Models: An Empirical Study and Beyond}

\author{Pengyu~Xue\orcidlink{0009-0007-3395-9575},
        Linhao~Wu\orcidlink{0009-0001-7624-156X},
        Zhongxing~Yu\orcidlink{0000-0003-3718-8476},\emph{Member, IEEE},
        Zhi~Jin\orcidlink{0000-0003-1087-226X},\emph{Fellow, IEEE},
        Zhen~Yang\orcidlink{0000-0003-0670-4538},\emph{Member, IEEE},
        Xinyi~Li\orcidlink{0009-0009-3728-5285},
        Zhenyu Yang\orcidlink{0000-0002-8334-7645},
        and Yue~Tan\orcidlink{0009-0007-2075-9398}

\thanks{Received 23 April 2024; revised 9 September 2024; accepted 6 October 2024. Recommended for acceptance by A. M. Moreno. This work was supported in part by the National Natural Science Foundation of China under Grant 62192731 and Grant 62102233, in part by the Natural Science Foundation of Shandong Province under Grant ZR2024QF093, and in part by Shandong Province Overseas Outstanding Youth Fund under Grant 2022HWYQ-043. \textit{(Pengyu Xue and Linhao Wu are co-first authors.) (Corresponding author: Zhen Yang.)}}
\thanks{Pengyu Xue, Linhao Wu, Zhongxing Yu, Zhen Yang, Zhenyu Yang, and Yue Tan are with the School of Computer Science and Technology, Shandong University, Qingdao, 266237, China. (e-mail: xuepengyu@mail.sdu.edu.cn, wulinhao@mail.sdu.edu.cn, yangzycs@mail.sdu.edu.cn, tanyue@mail.sdu.edu.cn, zhongxing.yu@sdu.edu.cn, zhenyang@sdu.edu.cn).} 
\thanks{Zhi Jin is with the Key Laboratory of High Confidence Software Technologies (Peking University), the Ministry of Education, Beijing, 100871, China, and also with the School of Computer Science, Peking University, Beijing, 100871, China. (e-mail: zhijin@pku.edu.cn).} 
\thanks{Xinyi Li is with the School of Electrical and Electronic Engineering, Nanyang Technological University, 639798, Singapore. (e-mail: Li\_XinYi@mail.sdu.edu.cn).}
\thanks{Digital Object Identifier 10.1109/TSE.2024.3478317}
}

\markboth{IEEE TRANSACTIONS ON SOFTWARE ENGINEERING}%
{XUE \MakeLowercase{et al.}: Automated Commit Message Generation with Large Language Models: An Empirical Study and Beyond}

\pagestyle{headings}

\maketitle

\begin{abstract}
Commit Message Generation (CMG) approaches aim to automatically generate commit messages based on given code \textit{diff}s, which facilitate collaboration among developers and play a critical role in Open-Source Software (OSS). Very recently, Large Language Models (LLMs) have been applied in diverse code-related tasks owing to their powerful generality. Yet, in the CMG field, few studies systematically explored their effectiveness. This paper conducts the first comprehensive experiment to investigate how far we have been in applying LLM to generate high-quality commit messages and how to go further beyond in this field.
Motivated by a pilot analysis, we first construct a multi-lingual high-quality CMG test set following practitioners' criteria.  
Afterward, we re-evaluate diverse CMG approaches and make comparisons with recent LLMs. To delve deeper into LLMs' ability, we further propose four manual metrics following the practice of OSS, including Accuracy, Integrity, Readability, and Applicability for assessment. Results reveal that LLMs have outperformed existing CMG approaches overall, and different LLMs carry different advantages, where GPT-3.5 performs best.

To further boost LLMs' performance in the CMG task, we propose an Efficient Retrieval-based In-Context Learning (ICL) framework, namely ERICommiter, which leverages a two-step filtering to accelerate the retrieval efficiency and introduces semantic/lexical-based retrieval algorithm to construct the ICL examples, thereby guiding the generation of high-quality commit messages with LLMs. Extensive experiments demonstrate the substantial performance improvement of ERICommiter on various LLMs across different programming languages. Meanwhile, ERICommiter also significantly reduces the retrieval time while keeping almost the same performance. Our research contributes to the understanding of LLMs' capabilities in the CMG field and provides valuable insights for practitioners seeking to leverage these tools in their workflows.
\end{abstract}


\section{Introduction}

\IEEEPARstart{I}{n} the domain of software development, well-structured commit messages are essential for they provide insights into related information of code changes, facilitating better understanding and collaboration among team members. However, writing commit messages manually can be time-consuming, and they are sometimes uninformative or even absent \cite{linares2015changescribe}. To overcome this issue, there have been substantial efforts in developing techniques to automatically generate commit messages, such as retrieval-based \cite{shi2022race,liu2018neural,liu2020atom}, and learning-based approaches \cite{tao2021evaluation,liu2019generating,lopes2024commit}.

In recent years, Large Language Models (LLMs) have shown exceptional performance across various domains of code intelligence, such as comment generation \cite{cai2023automated,yang2023significance}, code translation \cite{pan2023understanding,yang2024exploring}, and automated program repair \cite{xia2023automated}. Moreover, some studies have also conducted preliminary investigations on the performance of LLM in the CMG field. Nonetheless, they either only tested on limited LLMs \cite{lopes2024commit,zhang2024using}, such as ChatGPT, or merely included few and monolingual experimental samples\cite{zhang2024automatic}, lacking a systematic and comprehensive evaluation. Another study \cite{wang2023delving} focused on improving pre-trained language models' CMG performance by including commit-related issues as extra information. 
However, recent studies \cite{bachmann2010missing, mazrae2021automated, ruan2019deeplink, li2023commit} revealed that most of the commits (62.9\%) are not linked to issues, such as a bug fix or feature request, where false issue-commit links and less informative issues are prevalent. In addition, parameter tuning is also required, making it hard to apply to most LLMs in practice.

To fill the above gap, this paper conducts a systematic empirical study on diverse LLMs via automatic and manual evaluation, as well as proposes a training-free approach to boost their performance in the CMG task further. Initially, a pilot analysis in this paper investigated the data quality of the most widely used CMG dataset, namely MCMD \cite{tao2021evaluation}, and empirically proved its commit messages are of poor quality, lacking critical ``What'' (i.e., the specific changes made to the code) and ``Why'' (i.e., the reason or purpose behind code changes) information, which is deemed two essential elements for a good commit message \cite{tian2022makes}.
One of the most critical reasons is those commit messages were directly crawled from Version Control Systems (VCSs), where developers normally have little motivation and time to write informative commit messages \cite{tan2019communicate,maalej2010can}.
Considering most previous proposed CMG approaches \cite{shi2022race, jung2021commitbert, wang2021codet5} were assessed on datasets adopting almost the same workaround to extract CMG samples in the wild as MCMD, misleading guidance for developers was left in practice. 
Thus, we first construct a multi-lingual high-quality CMG test set based on the data cleaning and extracting from MCMD and another latest CMG dataset (i.e., CommitChronicle \cite{eliseeva2023commit}). This test set, dubbed MCMDEval+, strictly follows the requirements of practitioners and ensures all samples contain both ``What'' and ``Why'' information.
Subsequently, we further re-evaluate a wide range of state-of-the-art CMG approaches of various categories against recent LLMs, including GPT-3.5, LLaMA-7B/13B, and Gemini, in terms of a series of automated evaluation metrics following previous studies \cite{shi2022race,liu2024delving,liu2020atom,wang2021context}, showing that LLMs demonstrate superior performance against previous proposed CMG approaches overall. In particular, GPT-3.5 performs best and surpasses the best-performing CMG approach, namely RACE, by 78.79\% and 31.42\% in the METEOR and BLEU metrics, respectively. Notably, retrieval-augmented approaches perform better against other existing CMG approaches overall. In particular, NNGen, a purely retrieval-based CMG approach, performs the second best on average on MCMDEval+, demonstrating the importance of similar examples in the CMG task.     
To delve deeper into the CMG capability of LLMs, aligning with the requirements of OSS practice, we further conduct extensive manual assessments, covering aspects of Accuracy, Integrity, Readability, and Applicability. Manual assessments demonstrate that GPT-3.5 still performs the best among its various counterparts, but different LLMs carry different advantages. LLMs tend to convey more about ``What'' information but fewer details concerning ``Why'' owing to the limited available code context.

To better tap into the potential of LLMs and enhance the quality of commit messages generated by them, considering the impressive enhancement of ICL with LLM in extensive domains \cite{dong2022survey,bhattamishra2023understanding,wang2023survey,ovadia2023fine,li2024simac}, and the high impact of similar examples in the CMG task as mentioned before, we propose an Efficient Retrieval-based In-context learning framework (ERICommiter) applicable to various LLMs in the CMG task. ERICommiter leverages two-step filtering to extract high-quality example candidates to reduce the volume of the retrieval database, thereby accelerating retrieval efficiency. Subsequently, ERICommiter retrieves a group of the most similar examples for ICL, where both semantic and lexical retrieval engines are examined. 
Our experimental results demonstrate notable improvements in different metrics compared to using LLMs alone. Specifically, GPT-3.5 obtains improvements of 15.26\%, 4.75\%, and 13.72\% at most in terms of METEOR, BLEU, and ROUGE-L, respectively. Moreover, ERICommiter boosts the CMG performance of Gemini by 199.43\%, 155.36\%, and 167.77\% at most in terms of each evaluation metric in order. Besides, our proposed two-step filtering method reduces the retrieval time to 6.06\% for lexical-based retrieval and 7.16\% for semantic-based retrieval, concurrently keeping almost the same performance, which effectively improves the practicality of the LLM-based CMG system. 
This research contributes significantly to the realm of CMG, summarized as follows:

\begin{itemize}
    \item We construct a high-quality, multi-lingual CMG test set, namely MCMDEval+, to meet the practitioner's needs.
    \item We carry out the first systematic empirical study to investigate the performance of various recent LLMs against state-of-the-art CMG approaches of diverse categories. Besides, manual assessment aligning with the practical utility from a wide range of perspectives is proposed in this paper and conducted on LLM-generated commit messages. 
    \item We propose an Efficient Retrieval-based In-context learning framework applicable to LLMs in the CMG field, namely ERICommiter. Extensive experiments are conducted to demonstrate its effectiveness. 
    \item To facilitate future work, we release our replication package, including the cleaned dataset and the source code at \cite{Pengyu0324:online}
\end{itemize}

\section{Related Works}

\subsection{Commit Message Generation}
CMG aims to produce appropriate messages to describe a commit of code changes (i.e., code \textit{diff}s) in version control systems, which greatly facilitates collaboration among developers in Open-Source Software (OSS) practice. Over the years, various automated CMG approaches have been successively proposed.  
Liu et al. \cite{liu2018neural} propose a simpler and faster approach, named NNGen, to generate concise commit messages using the nearest neighbor algorithm. Tao et al. \cite{tao2021evaluation} perform a human evaluation and find the BLEU metric that best correlates with the human scores for the task. Liu et al. \cite{liu2020atom} propose ATOM, which advances commit messages by explicitly incorporating abstract syntax trees to represent code changes. Moreover, ATOM integrates both retrieved and generated commit messages using hybrid ranking. On top of that, the approach \cite{6975661} coined as ChangeScribe is proposed to generate commit messages automatically from change sets. RACE is a method that treats retrieved similar commits as an exemplar and utilizes it to generate readable and informative commit messages \cite{shi2022race}.
Recently, introducing retrieved relevant results into the training process has been found useful in most generation tasks \cite{zhan2021optimizing,gao2004dependence,berger2017information}.
Shi et al. \cite{shi2022race} treat the retrieved similar commit as an exemplar and train the model to utilize the exemplar for enhancing CMG.  Additionally, many neural-based approaches have been used to learn the semantics of code \textit{diff}s and translate them into commit messages. For example, NMTGen \cite{loyola2017neural} and Commit-Gen \cite{jiang2017automatically} adopt the Seq2Seq neural network with different attention mechanisms for translating them into commit messages. CommitBERT \cite{jung2021commitbert} leverages CodeBERT as an initial model to resolve the large gap between programming language and natural language, making it easier for the Neural Machine Translation (NMT)-based model to learn the contextual representation.  PtrGNCMsg \cite{liu2019generating} outperforms recent approaches based on the NMT structure, and first enables the prediction of out-of-vocabulary words. 
Very recently, researchers started to explore the potential of LLMs in the CMG field. Zhang et al. \cite{zhang2024automatic} explored the CMG performance of UniXcoder \cite{guo2022unixcoder} and ChatGPT \footnote{https://openai.com/blog/chatgpt/} for long code \textit{diff}s and whole message generation. Lopes et al. \cite{lopes2024commit} carried out a series of automatic and manual analyses on ChatGPT to investigate its CMG performance against previous proposed CMG approaches. However, different from this work, they either only experiment on limited LLMs or lack practical improvement and systematic evaluations for LLMs. 
\vspace{-0.2cm}
\subsection{What Is A Good Commit Message}

It is undeniable that the inconsistent and often incomplete nature of commit messages currently hinders effective communication and collaboration in OSS projects. Developers frequently write commit messages that vary in quality, as they often lack the time or motivation to ensure thoroughness. This inconsistency underscores the urgent need to establish a universal standard for what constitutes a good commit message, ensuring clarity and effectiveness in OSS collaboration.
Ma et al. \cite{ma2023improving} suppose that the quality of the commit message is in terms of informativeness, clearness, and length. On the other hand, Tian et al. \cite{tian2022makes} contend that the most frequently recognized expectation of a commit message is to summarize the changes in this commit ( noted as ``What'' ) and describe the reasons for the changes ( noted as ``Why'') through a survey of both academic papers and developer forums and validate the standards with experienced OSS developers. On top of that, they propose three classification models that can automatically identify and construct a high-quality commit message dataset. Additionally, some researchers have conducted studies based on Tian's understanding of what constitutes a good commit message. For instance, Li et al. \cite{li2023commit} cleaned Tian et al.'s dataset for good commit message identification by considering both the commit messages and the contents of the pull request/issue report.
Subsequently, they improved Tian et al's machine learning classifier to automatically identify whether a commit message contains ``What'' and ``Why'' information. In summary, it is commonly acknowledged \cite{tian2022makes, li2023commit, ma2023improving} that a good commit message should explain what was changed, and why a change was made. 

\vspace{-0.3cm}

\subsection{Large Language Models on Code}
Recently, a number of LLMs that are pre-trained on source code have been proposed, which mainly consist of three categories from the perspective of model structure, i.e., encoder-only models, decoder-only models, and encoder-decoder models.

\textbf{(1) Encoder-only models} contain an encoder only and are normally pre-trained with a series of code comprehension tasks, such as masked language modeling \cite{kenton2019bert} and replaced token detection \cite{clark2020electra}, leading to their powerful capability in code representation. Typical examples include CodeBERT \cite{feng2020codebert} and GraphCodeBERT \cite{guo2020graphcodebert}.
\textbf{(2) Decoder-only models} are pre-trained with the objective of next-token prediction language modeling in an unsupervised fashion \cite{li2023towards,li2023skcoder}, where GPT \cite{radford2018improving, radford2019language, brown2020language} series in the Natural Language Processing (NLP) field are successful paradigms. To this end, many decoder-only models optimized for code have been proposed based on the similar idea, such as LLaMA \cite{touvron2023llama}, CodeX \cite{chen2021evaluating}, GPT-CC \cite{CodedotA17:online}, and CodeGen \cite{nijkamp2022codegen}, which can be utilized in generation tasks.
\textbf{(3) Encoder-decoder models} are composed of an encoder and a decoder, which are typically pre-trained with denoising-based tasks. For example, CodeT5 \cite{wang2021codet5} is pre-trained with tasks of identifier tagging, masked identifier prediction, and bimodal dual generation. UniXCoder \cite{guo2022unixcoder} and PLBART \cite{ahmad2021unified} are pre-trained with the denoising sequence-to-sequence modeling task. Due to their encoder-decoder structure, tasks of both representation and generation can be successfully applied.

\section{Pilot Analysis}

Previous studies \cite{maalej2010can,tian2022makes,dyer2013boa,tan2019communicate} have confirmed that the quality of manually written commit messages on Version Control Systems (VCS) is generally poor. However, mainstream commit message datasets, such as the MCMD, have not yet been verified for quality issues, nor has there been research on cleaning and curating a high-quality dataset. This pilot analysis aims to conduct a series of quantitative and qualitative studies based on sampling to examine whether the MCMD dataset impacts the objective evaluation of CMG models and whether it requires cleaning.

\vspace{-0.2cm}
\subsection{Experimental design}

\subsubsection{Data set preparation}

This study first employs a Multi-programming-language
Commit Message Dataset (MCMD) for experimentation, which is the most widely used and recognized dataset in the CMG field \cite{shi2022race,tao2021evaluation,liu2018neural}. The MCMD dataset covers five major Programming Languages (PLs), including Java, Python, JavaScript, C++, and C\#. Noting that each commit in MCMD corresponds to a single code \textit{diff} file. To control experimental costs while ensuring a statistically valuable evaluation, 
%we sample data points on the test set of MCMD with a 95\% statistical confidence and 5\% confidence interval method from PLs of Java, Python, and JavaScript. 
we sample data points from the MCMD test set for Java, Python, and JavaScript, using a 95\% statistical confidence and a 5\% confidence interval method, focusing our evaluation on these three prevalent PLs.
Detailed sample volumes of each PL are shown in Table \ref{num}. This design enhances the statistical significance of experimental results, strengthening the research's empirical foundation.

\begin{table}[htbp]
\setlength{\abovecaptionskip}{0cm}
\vspace{-0.4cm}
\footnotesize
\centering
\caption{Data Volume Before and After Sampling}
\label{num}
\renewcommand{\arraystretch}{0.8}
\begin{tabular}{ccc}
\toprule
PL& Test Sets Volume& Sampling Volume\\
\midrule
Java&20159&376\\
Python&25837&378\\
JavaScript&24773&378\\
\bottomrule
\end{tabular}
\vspace{-0.2cm}
\end{table}

\subsubsection{Experimental Model Preparation}
\label{Experimental Model Preparation}
In this preliminary experiment, we utilize a state-of-the-art LLM, i.e., GPT-3.5, for experiments. 
GPT-3.5 is a decoder-only LLM, pre-trained on a huge amount of human-written text/code, capable of capturing complex language patterns and contextual information. 
In the experiment, we adopt the version of gpt-3.5-turbo and set the following model parameters: \textit{max\_tokens=}50 for controlling the length of generated messages. \textit{temperature=}0.8 and \textit{top\_p=}0.95 by default, respectively. Additionally, all other parameters are kept at their default settings.
Following previous studies \cite{lopes2024commit, zhang2024using}, we design a basic prompt for GPT-3.5 with the most simple instructions and a zero-shot setting to mimic practical usage. The basic prompt can be formally defined as:
\textit{``\$\{Code\_Diff\}\textbackslash nYou are a programmer who makes the above code changes. Please write a commit message for the above code change.''}, where \textit{\$\{Code\_Diff\}} is the placeholder for a code \textit{diff} snippet. Afterward, we obtain its first output for evaluation. 
The generated results are collected and compared with those produced by the state-of-the-art CMG approach, i.e., RACE\cite{shi2022race} in terms of automated assessment metrics mentioned in Section \ref{Automated assessment}.

\subsubsection{Automated assessment}
\label{Automated assessment}

Following the previous studies in the CMG field \cite{shi2022race,lopes2024commit,zhang2024automatic}, we adopt METEOR, BLEU, and ROUGE-L to assess the performance of each model.
\begin{itemize}
    \item METEOR\cite{banerjee2005meteor}: This metric measures the word-level matching between machine-generated text and reference text. METEOR considers several factors, including exact word matches, synonym matches, and stemming (morphological variations).   It also accounts for word order by incorporating a penalty for misaligned words.
    \item BLEU\cite{papineni2002bleu}: The computation of this metric relies on the  overlap of n-grams (contiguous sequences of n words) between the generated text and one or more reference texts.
    \item ROUGE-L\cite{lin2004rouge}: This metric focuses on the longest common subsequence (LCS) between the generated text and the reference text, which measures the longest sequence of words that appears in both texts in the same order. 
\end{itemize}

\vspace{-0.2cm}
\subsection{Experimental results and analysis}

Comparing the results between GPT-3.5 and RACE on the sampled dataset, we find that GPT-3.5 does not outperform RACE regarding automated assessment metrics as shown in Table \ref{preexp}, which contradicts the usual impression reflected in the literature concerning LLM-driven code intelligence \cite{xia2023automated,chang2023survey,pan2023understanding}.  
To delve deeper into the reasons behind this abnormal result, we introduce manual evaluation. Specifically, we select commit messages generated by GPT-3.5 with BLEU scores lower than 50\% of those generated by RACE for human analysis and strictly follow the criterion of good commit messages \cite{tian2022makes} to proceed with the assessment, thereby exploring why GPT-3.5 does not perform as well as we expected.

\begin{table}[!ht]
\vspace{-0.3cm}
\setlength{\abovecaptionskip}{0cm}
\footnotesize
\centering
\caption{Performance Comparison between GPT-3.5 and RACE on the Sampled Dataset}
\setlength{\tabcolsep}{3.4pt} 
\label{preexp}
    \centering
    \resizebox{\linewidth}{!}{
    \begin{tabular}{cccccccccc}
    
    \toprule
        Model & ~ & Java & ~ & ~ & Python & ~ & ~ & JS & ~ \\ 
        \midrule
        ~ & BLEU & Met. & Rou. & BLEU & Met. & Rou. & BLEU & Met. & Rou. \\ 
        
        RACE & \bf{25.66} & \bf{15.46} & \bf{32.02} & \bf{21.79} & \bf{14.68} & \bf{28.35} & \bf{25.55} & \bf{16.31} & \bf{31.79} \\ 
        GPT-3.5 & 18.94 & 12.79 & 15.27 & 17.85 & 8.88 & 11.36 & 15.56 & 8.31 & 12.00 \\ 
    \bottomrule
    \end{tabular}
    }
\vspace{-0.3cm}
\end{table}

According to the widely acknowledged definition in previous studies\cite{wang2023delving,li2023commit,tian2022makes},
a good commit message contains two key elements of ``What'' (i.e., what was done) and ``Why'' (i.e., why it was done). The ``What'' part reveals the specific changes made in the code, such as fixing a bug, adding a new feature, and improving existing code. The ``Why'' section provides background information on the changes, explaining why such modifications are necessary, such as enhancing performance, addressing security vulnerabilities, or improving user experience.

During the manual assessment, we select the results of GPT-3.5 and ground truths for comparison, thereby analyzing the specific areas where GPT-3.5 falls short.
To ensure a thorough analysis, we use a structured approach where six authors with 3-5 years of development experience in this article serve as evaluators, and each pair is responsible for assessing one PL. Specifically, each evaluator is assigned a set of ground truth and GPT-3.5-generated commit message pairs for assessment. To ensure the accuracy and reliability of the assessment results, we adopt a dual assessment mechanism \cite{bryman2016social} whereby two independent evaluators assess each commit message together. Subsequently, we use Cohen’s kappa coefficient \cite{chmura2002kappa} to measure the agreement of the assessment results between the two evaluators, a high kappa score indicating a higher level of reliability in the evaluation results.
For each ground truth and GPT-3.5-generated commit message sample, evaluators assess the following two aspects:

\textbf{Element 1: Existence of ``What'' Content (0/1):}
Check whether the generated statements clearly present a description or expression of what has been done.

\textbf{Element 2: Existence of ``Why'' Content (0/1):}
Examine whether the generated text provides the reasons or background for why this was done.

\begin{figure}[!t]
\vspace{-1.5em}
\setlength{\abovecaptionskip}{0cm}
  \centering
  \includegraphics[width=\linewidth]{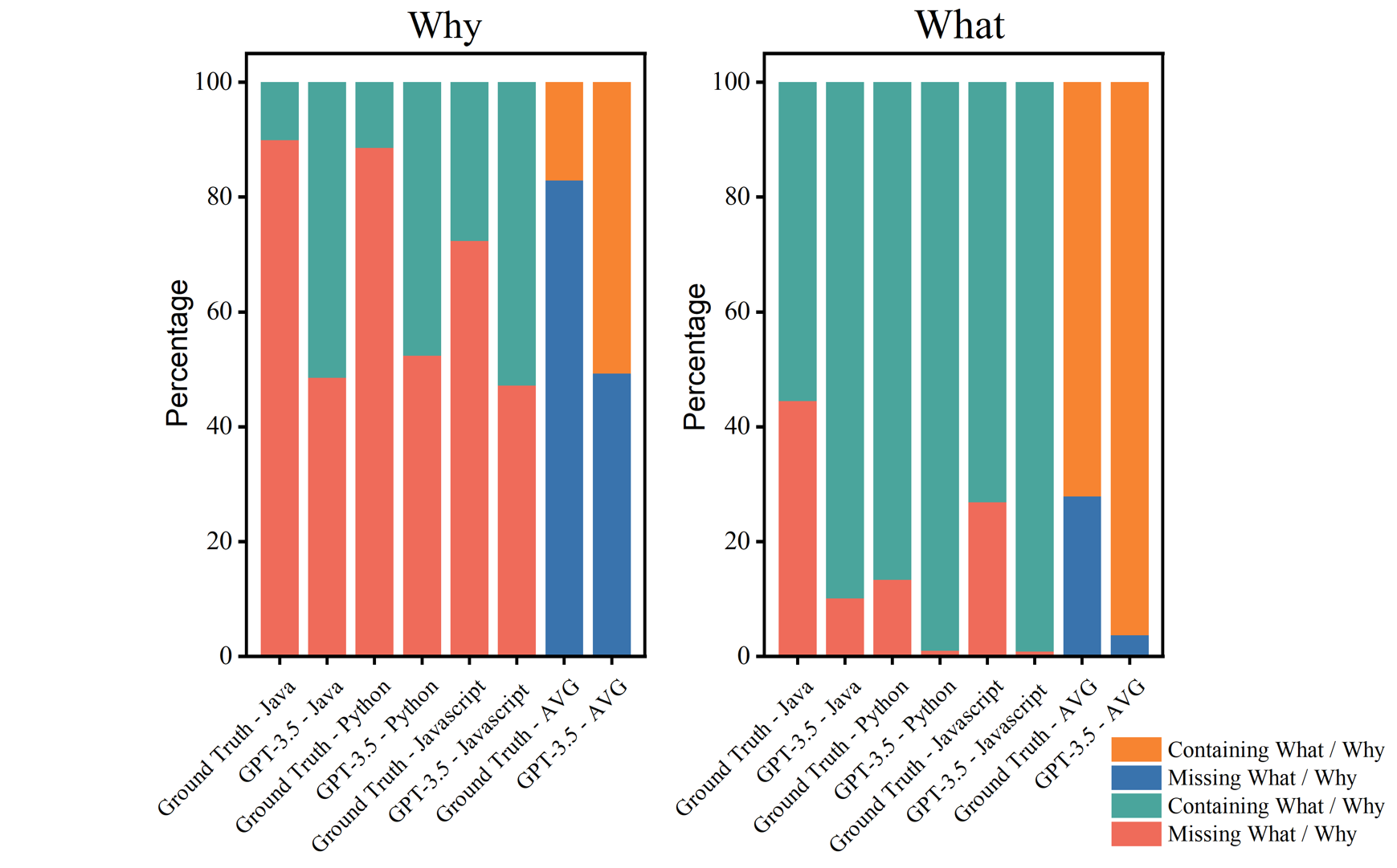}
  \caption{The Percentage (\%) of Containing/Missing ``What''/``Why'' Elements in Comparison between GPT-3.5-Generated Commit Messages and Ground Truths.}
    \label{missing}
\vspace{-2em}
\end{figure}

In the manual assessment, the Kappa score of the assessment results is 0.65, indicating a high degree of consistency between the evaluators, thus demonstrating the accuracy and reliability of our assessment method.
The specific evaluation results are shown in Fig.~\ref{missing}. where each bar denotes a comparison between GPT-3.5-generated commit messages and ground truths in terms of the percentage of their commit messages containing or missing ``What''/``Why'' information. For example, GPT-3.5-Java represents the evaluation result of commit messages for the Java PL generated by GPT-3.5.

The research results indicate that some ground truth commit messages themselves are not entirely accurate or detailed. Many ground truth messages lack descriptions of ``What'' and ``Why'', while the GPT-3.5 demonstrates a relatively superior ability to generate these contents. This implies that from the perspectives of practitioners, GPT-3.5-generated commit messages are more informative and of higher quality compared with ground truths. 
We conjecture that the ground truths of the MCMD dataset are inherently flawed, as their samples are extracted from VCSs, such as Github, in the wild, where developers normally lack the motivation, time, and experience to write high-quality commit messages \cite{tan2019communicate,maalej2010can}.
Considering most of the previous CMG approaches are evaluated on the MCMD dataset \cite{shi2022race,liu2018neural,loyola2017neural,jiang2017automatically,wang2021context,liu2019generating}, and other related CMG datasets may also exhibit the aforementioned flaws, as crawling samples from VCSs are a common workaround in the CMG field,
it is urgent to clean these datasets up and reassess previous mainstream CMG approaches, including LLMs.

\begin{boxK}
  \faIcon{pencil-alt}  \textbf{Conclusion of Pilot Analysis:}
  
  (1) The MCMD dataset, widely used in the CMG tasks, can not accurately reflect the current capabilities of the CMG approaches due to the low data quality. 
  
  (2) There is an urgent need for a thorough cleanup of CMG-related datasets (e.g., MCMD) and reassessment to ensure more objective and precise evaluations of the CMG techniques in the following sections.
\end{boxK}

\vspace{-0.3cm}

\section{Construction of the high-quality test set}
\label{Construction of the high-quality test set}
As mentioned above, the quality of ground truths significantly affects the objectivity and authenticity of assessment. Therefore, a pivotal step in this study is to construct a meticulously curated, higher-quality, and reliable CMG dataset covering mainstream PLs for assessing and validating the CMG performance of diverse models systematically.
To ensure the comprehensiveness of PLs, we introduce another dataset, CommitChronicle \cite{eliseeva2023commit}, which contains 10.7M commits across 20 PLs. Based on the datasets of MCMD and CommitChronicle, we construct the MCMD+ dataset (refer to Table A in \cite{Pengyu0324:online}) for our experiments. This dataset encompasses five PLs (including Java, Python, JavaScript, C++, and C\#) from MCMD and three PLs (including
Go, PHP, and Rust) from CommitChronicle. In CommitChronicle, a single commit may correspond to multiple \textit{diff} files. To maintain consistency in the MCMD+ dataset, we retain only the commit data associated with a single \textit{diff} file from CommitChronicle. Additionally, we convert the format of the whole data in CommitChronicle to match that of MCMD. Considering the trade-off between limited human effort and assessment validity, we meticulously curate 500 high-quality samples from each selected PL of the test set of MCMD+, thereby constructing a brand-new test set, namely MCMDEval+ in this work.
Throughout this process, to improve the efficiency of sample selection as much as possible, we carry out a meticulous three-stage screening on MCMD+, including (1) filtering out non-target languages, (2) automatically filtering good commit messages through a deep learning model, and finally (3) further selecting high-quality samples manually.
We hope the newly constructed test set can not only be used to evaluate the performance of CMG models but also serve as reference templates for developers to enhance code quality and collaboration efficiency within teams. 

\textbf{Step 1: Precise Filtering of Non-Target PL samples.}
Based on our observations, although the data from MCMD is divided into five categories according to PLs, each sub-dataset of a PL is mixed with a large amount of data from other PLs. For instance, when dealing with a sub-dataset targeting Java, we encounter the inclusion of many code \textit{diff}s from other PLs, such as C++, Python, and even HTML. This issue arises because its PL-specific sub-datasets are roughly partitioned according to the major PL of projects that contain multiple minor PLs according to the explanation of the original authors. Thus, this severely affects the accuracy of previous studies' experimental results under different PLs and poses a major obstacle to the evaluation in this paper. 

To tackle this problem, we employ regular expressions as the primary filtering tool, where files with suffixes that do not belong to the target PL (i.e., .java, .py, .js, .cpp, .cs) are filtered out, thereby completing the first-step cleaning. It is worth noting that in the CommitChronicle dataset released by Aleksandra et al.  \cite{eliseeva2023commit}, all PLs are mixed in a single file, without any subdivision by PL. Therefore, the original data volume per PL is represented by ``-'' in Table \ref{exp5}. Consequently, we also employ regular expressions on file suffixes to extract samples of selected PLs (i.e., .go, .php, .rs) from CommitChronicle, ensuring the PL purity of each sub-dataset in this step. 
Eventually, each PL-specific sub-dataset in MCMD+ only contains samples of their targeted PLs, and their data volume after this first-step cleaning is shown in Table \ref{exp5}.

\begin{table}[]
\setlength{\abovecaptionskip}{0cm}
\footnotesize
\centering
\caption{Data Volume Statistics After Filtering at Each Step}
\label{exp5}
\renewcommand{\arraystretch}{1.2}
\resizebox{\linewidth}{!}{
\begin{tabular}{cccccc|c}
\toprule
Test Set                 & PL         & Original & Step1 & Step2 & Step3 & Kappa \\ \hline
\multirow{5}{*}{MCMD}            & Java       & 20159    & 11746 & 4167  & 500   & 0.59  \\
                                 & Python     & 25837    & 14012 & 4601  & 500   & 0.56  \\
                                 & C\#        & 18702    & 9224  & 3432  & 500   & 0.62  \\
                                 & C++        & 20141    & 5778  & 2029  & 500   & 0.57  \\
                                 & JavaScript & 24773    & 10816 & 3855  & 500   & 0.61  \\ \hline
\multirow{3}{*}{CommitChronicle} & Go         & -        & 33838 & 8954  & 500   & 0.58  \\
                                 & PHP        & -        & 9788  & 3229  & 500   & 0.60  \\
                                 & Rust       & -        & 13588 & 4550  & 500   & 0.61  \\ \bottomrule
\end{tabular}
}
\vspace{-0.2cm}
\end{table}

\textbf{Step 2: Automated Filtering of samples Containing ``What'' and ``Why''.}

In our study, we recognize that manually filtering out commit messages to identify those containing both ``What'' (what was done) and ``Why'' (why it was done) elements is time-consuming and labor-intensive. To address this issue, we introduce an automated method based on Bi-LSTM \cite{zhang2015bidirectional} to efficiently identify high-quality commit messages, which was proposed by \cite{tian2022makes} and adhered to the definition of ``good commit message'' during the identification.

Since Tian et al. \cite{tian2022makes} did not publish the trained parameters, we try our best to replicate this Bi-LSTM model following their instructions and train the model from scratch. Afterward, we apply it to our commit messages following their preprocessing procedures such as text cleaning, vocabulary standardization, and encoding.
The replication results are shown in Table \ref{performance}, where ``C-Why'', ``C-What'', and ``C-Good'' represent the performance of Bi-LSTM on identifying commit messages containing ``Why'', ``What'', and both elements. Although the replicated performance is slightly weaker than that reported in the original paper \cite{tian2022makes}, it still effectively filters out those high-quality commit messages to a great extent. After the filtering process, approximately one-third of the data remained from step-1 filtering, as shown in Table \ref{exp5}, which further improves the density of high-quality samples and is helpful in reducing the efforts of manual selection in the third step.

\begin{table}[]
\setlength{\abovecaptionskip}{0cm}
\vspace{-1em}
\footnotesize
\centering
\caption{ Performance of Bi-LSTM in Our Study }
\label{performance}
\renewcommand{\arraystretch}{1.2}
\begin{tabular}{cccc}
\toprule
Metrics                       & C-Why                       & C-What                      & C-Good                      \\ \hline
\multicolumn{1}{l}{Precision} & \multicolumn{1}{l}{44.60\%} & \multicolumn{1}{l}{85.37\%} & \multicolumn{1}{l}{68.56\%} \\
Recall                        & 52.03\%                     & 34.80\%                     & 51.34\%                     \\
F1                            & 48.03\%                     & 47.01\%                     & 58.47\%                     \\ \bottomrule
\end{tabular}
\vspace{-2em}
\end{table}

\textbf{Step 3: Manual Selection of Samples with ``What'' and ``Why''.}

After the automated filtering process, we proceed to the stage of manual assessment, aiming to select high-quality commit messages from the remaining samples precisely, where  
four authors of this article, all with 3-5 years of programming experience, serve as evaluators. 
Specifically, we employ a dual-evaluator mechanism, dividing evaluators into pairs and conducting independent evaluations in a back-to-back manner \cite{bryman2016social} to minimize the mutual influence on opinions. Each evaluator assigns a score of 0 or 1 to each commit message. Here, a score of 0 indicates that the commit lacks either the ``What'' or ``Why'' element, while a score of 1 indicates that the commit contains a clear and detailed description of ``What'' and ``Why''. When both evaluators provide consistent scores (either both 0 or both 1), deciding whether to retain or discard the commit is simple: discard for all 0s or retain for all 1s. In instances of conflicting scores, a third evaluator will be brought in for arbitration.
To validate the effectiveness of the assessment, we compute Cohen's kappa coefficient \cite{chmura2002kappa} of agreement for the scores assigned by the two evaluators. 
The kappa scores for each language are calculated as shown in Table \ref{exp5}. In our case, the overall mean value of the kappa score is 0.59, which shows that our assessment process is effective. Once the number of manually selected samples of each PL reaches 500, we stop the above process of this step.

Through such screening, we ultimately construct a high-quality, multi-lingual test set comprising 500 CMG samples with comprehensive information (e.g., code \textit{diff}, commit messages, and rich metadata as MCMD) for each of the eight PLs, we name it MCMDEval+. 
To further verify the quality of the constructed test set, we conduct a third-party survey. Specifically, we randomly select 10 samples from MCMDEval+ for each PL, with each sample corresponding to a code \textit{diff} and a commit message. Subsequently, we design a questionnaire comprising the 80 extracted samples, organized into eight sections by PL. We then invite two experts with extensive coding experience to manually evaluate their quality. The experts are asked to independently rate the quality of each commit message corresponding to a code \textit{diff} on a 5-point Likert scale (from “very poor” 1 to “very good” 5) \cite{bertram2007likert} based on the definition of a good commit message and their own practical experience. Finally, we calculate the average score for all samples to be 4.25, demonstrating a quite high recognition of the samples’ commit message quality. To validate the effectiveness of the evaluation, we compute the Cohen’s kappa coefficient \cite{chmura2002kappa} to assess the inter-rater agreement between the two experts, obtaining a score of 0.67, indicating that they have reached a ``substantial agreement''. Consequently, we believe that the verification of MCMDEval+ is valid. 
We have open-sourced MCMDEval+ at \cite{Pengyu0324:online}. 

\vspace{-0.3cm}

\section{Study Design}

This section details the experimental design, including studied models, implementation, and evaluation methodology.
\vspace{-0.3cm}

\subsection{Models and Implementations}

In this study, we select the most widely used and state-of-the-art CMG approaches for a systematic comparison with a series of recent LLMs of diverse sizes and families.

\textbf{NNGen \cite{liu2018neural}:} 
A retrieval-based approach employs the nearest neighbor algorithm to fetch the top-k similar code \textit{diff}s. NNGen selects the most similar result based on BLEU scores to generate commit messages.

\textbf{CommitGen \cite{jiang2017automatically}:} A model that treats code \textit{diff}s as plain texts and adopts a Seq2Seq neural network with different attention mechanisms to translate them into commit messages.

\textbf{NMTGen \cite{loyola2017neural}:} In this context, NMTGen is used to learn the semantics of code \textit{diff}s and translate them into commit messages.

\textbf{PTrGNCMSG \cite{liu2019generating}:} A model that incorporates the pointer-generator network into the Seq2Seq model to handle out-of-vocabulary words.

\textbf{CoRec \cite{wang2021context}:} 
A hybrid model employing an encoder-decoder architecture that incorporates retrieved results during inference. The model encode the input and retrieved code \textit{diff}, combining the probability distributions of each word to generate commit messages.

\textbf{CodeBERT \cite{feng2020codebert}:} Leverages CodeBERT, a pre-trained language model for source code, to learn the semantic representations of code \textit{diff}s and adopts a Transformer-based decoder to generate commit messages.

\textbf{RACE \cite{shi2022race}:} A retrieval-augmented neural CMG method that achieved the state-of-art performance. RACE enhances commit message accuracy by using a similar retrieved commit as an exemplar, with an integrated exemplar guider that aligns the retrieved and current code \textit{diff} and then guides the generation of commit message based on semantic similarity.

\textbf{LLaMA \cite{touvron2023llama}:} 
A family of multilingual large language models employing a decoder-only transformer architecture offers configurations ranging from 7B to 65B in parameter size. LLaMA is pre-trained from various open-source datasets of up to 1.4 trillion tokens. We include two versions of 7B/13B for experiments and use APIs provided by HuggingFace for implementation.

\textbf{Gemini \cite{team2023gemini}:} 
A family of multimodal models which build on top of Transformer decoders. Gemini models are trained on a dataset that is both multimodal and multilingual.  Gemini family consists of Ultra, Pro, and Nano three sizes. We employ the Gemini-Pro model and use Google's APIs for experiments.

\textbf{GPT-3.5 \cite{brown2020language}:} 
A model based on the GPT-3 architecture and fine-tuned with RLHF techniques, performing exceptionally well in various natural language processing and code related tasks.  
The same as the pilot analysis, we adopt the version of gpt-3.5-turbo for experiments.

For the implementation of LLMs, we adhere to the hyper-parameter configuration detailed in Section \ref{Experimental Model Preparation} and assess their performance on MCMDEval+. For the evaluation of other state-of-the-art CMG approaches, towards PLs from the MCMD dataset, we extract their generated commit messages corresponding to our cleaned 500 samples on Java, Python, C\#, C++, and JavaScript, respectively, for re-evaluation and comparative analysis, as 
corresponding experimental results have been released at \cite{tao2021evaluation}. In contrast, towards PLs from the CommitChronicle dataset, as cleaned samples of Go, PHP, and Rust have not been tested on our selected CMG approaches before, we re-implement and re-train them based on their published repositories on the training set of MCMD+ and assess their performance on these PLs of MCMDEval+. 
Furthermore, our experiments are conducted on a Windows 11 laptop, using Python 3.10, openai v1.35.9, scikit-learn 1.5.0, nltk 3.8, and rouge 1.0.0 for programming.
\vspace{-0.2cm}

\vspace{-0.12cm}

\subsection{Evaluation Methodology}
\label{Evaluation Methodology}
To achieve a systematic evaluation, we first adopt automated assessment in terms of METEOR, BLEU, and ROUGE-L, as mentioned in Section \ref{Automated assessment}, to make comparisons among current CMG approaches and recent LLMs. This solves the problem: \textbf{RQ1: How is the performance of recent LLMs against current CMG approaches?}

To delve deeper into the LLMs' performance in the CMG task, we further arrange a series of manual assessments and include the currently best-performing CMG approach, namely RACE, for further comparison. 
Aligning with the practice of OSS, we not only focus on the existence of ``What''/``Why'' elements (i.e., Integrity) for LLM/RACE-generated commit messages but also involve Accuracy, Readability, and Applicability in the assessment, as these aspects impact the internal and external quality of the software product and their usability in OSS practice \cite{ma2023improving, yilmaz2022quality}. Consequently, we propose the following research questions with the definitions of each human evaluation metric below.

\textbf{RQ2: How accurate is the commit message generated by LLMs?}
Different from the automated metric mentioned above, we define Accuracy as a manual metric that reflects more on semantic equivalency between model-generated commit messages and ground truths instead of their rigid literal consistency, focusing more on the deep contextual and semantic understanding that automated metrics may overlook.

\textbf{RQ3: How integral is the commit message generated by LLMs?}
We define Integrity as the extent to which a model-generated commit message contains both the ``What'' and ``Why'' elements, which are hard to be precisely detected in an automated manner. The above measurement is critical, as commit messages implying clear modification illustration (i.e., ``What'' element) and motivation (i.e., ``Why'' element) allow for an easier understanding of code changes, simplify the code review process, and aid debugging in OSS practice.

\textbf{RQ4: How readable is the commit message generated by LLMs?}
We define Readability as the clarity and understandability of commit messages. 
Readable commit messages enhance communication among team members, facilitate code maintenance, and contribute to effective project management.  
Evaluators are required to evaluate commit messages based on factors such as fluency and correct use of grammar. 

\textbf{RQ5: How applicable is the commit message generated by LLMs?}
We define Applicability as the willingness of OSS developers to adopt a model-generated commit message for their code changes. 
Unlike the previously mentioned three metrics, which each assess a specific aspect of commit message quality, Applicability is a more holistic metric. It incorporates the considerations of the aforementioned metrics into the actual decision-making and usage intentions of developers, placing greater emphasis on whether the generated commit message is adopted by OSS developers. This provides a more comprehensive evaluation perspective.

To address the research questions RQ2$\sim$RQ5,
We invite 16 full-time developers from our networks within the software industry, including 3 front-end developers specializing in JavaScript, 4 back-end developers proficient in Python, Java, and PHP, 3 system programmers experienced in C++ and Rust, 2 cloud and distributed systems engineers skilled in Go, and 4 full-stack developers with expertise in both front-end and back-end technologies, each with 3-5 years of experience, to serve as evaluators in the interviews.
Specifically, we employ the three-level Likert scale which allows evaluators to express their degrees of positive or negative attitudes based on their experience and observations, from ``disagree'', ``neutral'' to ``agree,'' or from ``poor'', ``neutral'' to ``good''\cite{bertram2007likert} in interviews for feedback collection. 
Based on the above, we design eight questionnaires corresponding to the eight PLs studied and each questionnaire is distributed to two evaluators based on their expertise. Every questionnaire comprises ten CMG samples, each exploring the performance of different LLMs and the currently best-performing CMG approach, i.e., RACE, in generating commit messages on various aspects, including Accuracy, Integrity, Readability, and Applicability.  
Finally, we obtain 16 interview results, covering eight PLs. Considering each evaluator is required to answer the full range of questions for each PL, in the validation of the effectiveness of this manual assessment,
we separately calculate Cohen's kappa consistency coefficients \cite{chmura2002kappa} between the scores given by the two experts.

Particularly, in order to understand individual behavior and decision-making in
the Applicability assessment, we use an ``immersive context'' approach, a technique commonly used in sociological research\cite{liu2019using}. 
Specifically, evaluators are guided to place themselves in a hypothetical role, assuming they are the programmers writing the evaluated information. We inquire with evaluators about their willingness to adopt the provided commit message if they have authored the code in question.
Through the process mentioned above, we are able to holistically assess the overall capabilities of LLMs in CMG, providing valuable insights and guidance for the field of software development.

\begin{table}[]
\setlength{\abovecaptionskip}{0cm}
\vspace{-1em}
\footnotesize
\centering
% \caption{Comparison of LLMs with other models under three metrics on five PLs.}
\caption{Performance Comparison of LLMs Against State-of-the-art CMG Approaches}
\label{nlpresult}
\renewcommand{\arraystretch}{1.2}
\resizebox{\linewidth}{!}{
\begin{tabular}{
>{\columncolor[HTML]{FFFFFF}}c 
>{\columncolor[HTML]{FFFFFF}}c 
>{\columncolor[HTML]{FFFFFF}}c 
>{\columncolor[HTML]{FFFFFF}}c 
>{\columncolor[HTML]{FFFFFF}}c 
>{\columncolor[HTML]{FFFFFF}}c 
>{\columncolor[HTML]{FFFFFF}}c 
>{\columncolor[HTML]{FFFFFF}}c 
>{\columncolor[HTML]{FFFFFF}}c 
>{\columncolor[HTML]{FFFFFF}}c 
>{\columncolor[HTML]{FFFFFF}}c 
>{\columncolor[HTML]{ECF4FF}}c }
\toprule
\multicolumn{2}{c}{\cellcolor[HTML]{FFFFFF}Model}                                                           & Metric & Java           & Python         & C\#            & C++            & JS             & Go             & PHP            & Rust           & Avg.                     \\ \hline
\cellcolor[HTML]{FFFFFF}                              & \cellcolor[HTML]{FFFFFF}                            & Met.   & 14.28          & 8.66           & 12.41          & 7.82           & 10.60          & 5.01           & 4.76           & 5.21           & 8.59                     \\
\cellcolor[HTML]{FFFFFF}                              & \cellcolor[HTML]{FFFFFF}                            & BLEU   & 22.23          & 20.58          & 21.54          & 16.53          & 19.84          & 6.89           & 7.32           & 7.06           & 15.25                    \\
\multirow{-3}{*}{\cellcolor[HTML]{FFFFFF}IR-based}    & \multirow{-3}{*}{\cellcolor[HTML]{FFFFFF}NNGen}     & Rou.   & 18.85          & 15.82          & 17.96          & 13.38          & 16.25          & 8.81           & 7.60           & 7.15           & 13.23                    \\ \hline
\cellcolor[HTML]{FFFFFF}                              & \cellcolor[HTML]{FFFFFF}                            & Met.   & 2.38           & 1.71           & 1.92           & 1.12           & 1.98           & 5.26           & 4.75           & 3.89           & 2.88                     \\
\cellcolor[HTML]{FFFFFF}                              & \cellcolor[HTML]{FFFFFF}                            & BLEU   & 3.77           & 4.01           & 3.87           & 2.91           & 3.83           & 1.93           & 1.88           & 1.77           & 3.00                     \\
\cellcolor[HTML]{FFFFFF}                              & \multirow{-3}{*}{\cellcolor[HTML]{FFFFFF}CommitGen} & Rou.   & 6.89           & 6.94           & 4.83           & 4.38           & 6.12           & 5.94           & 5.16           & 2.02           & 5.29                     \\
\cellcolor[HTML]{FFFFFF}                              & \cellcolor[HTML]{FFFFFF}                            & Met.   & 2.64           & 2.41           & 1.89           & 1.62           & 3.28           & 2.16           & 2.31           & 2.74           & 2.38                     \\
\cellcolor[HTML]{FFFFFF}                              & \cellcolor[HTML]{FFFFFF}                            & BLEU   & 3.93           & 4.84           & 3.36           & 3.37           & 6.48           & 3.34           & 4.25           & 3.72           & 4.16                     \\
\cellcolor[HTML]{FFFFFF}                              & \multirow{-3}{*}{\cellcolor[HTML]{FFFFFF}NMTGen}    & Rou.   & 7.30           & 8.35           & 4.86           & 5.00           & 6.96           & 8.18           & 7.83           & 7.34           & 6.98                     \\
\cellcolor[HTML]{FFFFFF}                              & \cellcolor[HTML]{FFFFFF}                            & Met.   & 11.86          & 7.48           & 8.57           & 6.8            & 8.05           & 7.39           & 7.69           & 8.14           & 8.25                     \\
\cellcolor[HTML]{FFFFFF}                              & \cellcolor[HTML]{FFFFFF}                            & BLEU   & 14.95          & 16.55          & 12.42          & 12.35          & 13.46          & 10.36          & 10.72          & 10.31          & 12.64                    \\
\multirow{-9}{*}{\cellcolor[HTML]{FFFFFF}End-to-end}  & \multirow{-3}{*}{\cellcolor[HTML]{FFFFFF}PTrGNCMSG} & Rou.   & 17.02          & 15.68          & 12.64          & 13.05          & 14.21          & 12.52          & 11.28          & 10.76          & 13.40                    \\ \hline
\cellcolor[HTML]{FFFFFF}                              & \cellcolor[HTML]{FFFFFF}                            & Met.   & 6.42           & 4.85           & 4.90           & 2.79           & 5.24           & 4.98           & 1.96           & 4.18           & 4.42                     \\
\cellcolor[HTML]{FFFFFF}                              & \cellcolor[HTML]{FFFFFF}                            & BLEU   & 10.30          & 10.49          & 8.41           & 6.58           & 9.66           & 5.18           & 3.03           & 7.30           & 7.62                     \\
\cellcolor[HTML]{FFFFFF}                              & \multirow{-3}{*}{\cellcolor[HTML]{FFFFFF}CoRec}     & Rou.   & 12.04          & 12.32          & 9.80           & 7.83           & 10.46          & 5.28           & 4.59           & 7.54           & 8.73                     \\
\cellcolor[HTML]{FFFFFF}                              & \cellcolor[HTML]{FFFFFF}                            & Met.   & 11.54          & 10.61          & 12.11          & 6.66           & 11.09          & 11.84          & 10.64          & 9.64           & 10.52                    \\
\cellcolor[HTML]{FFFFFF}                              & \cellcolor[HTML]{FFFFFF}                            & BLEU   & 21.41          & 23.75          & 22.74          & 16.01          & 21.41          & 15.16          & 18.74          & 19.97          & 19.90                    \\
\multirow{-6}{*}{\cellcolor[HTML]{FFFFFF}Hybrid}      & \multirow{-3}{*}{\cellcolor[HTML]{FFFFFF}RACE}      & Rou.   & \textbf{19.08} & \textbf{21.00} & \textbf{20.46} & 12.57          & \textbf{19.23} & \textbf{18.62} & \textbf{19.05} & 18.25          & \textbf{18.53}           \\ \hline
\cellcolor[HTML]{FFFFFF}                              & \cellcolor[HTML]{FFFFFF}                            & Met.   & 9.59           & 7.66           & 7.72           & 4.51           & 7.58           & 4.56           & 3.92           & 4.84           & 6.30                     \\
\cellcolor[HTML]{FFFFFF}                              & \cellcolor[HTML]{FFFFFF}                            & BLEU   & 14.01          & 16.56          & 12.46          & 11.23          & 13.12          & 10.63          & 6.12           & 10.71          & 11.86                    \\
\multirow{-3}{*}{\cellcolor[HTML]{FFFFFF}Pre-Trained} & \multirow{-3}{*}{\cellcolor[HTML]{FFFFFF}CodeBERT}  & Rou.   & 16.80          & 17.73          & 14.38          & 9.87           & 15.72          & 10.34          & 7.49           & 9.87           & 12.78                    \\ \hline
\cellcolor[HTML]{FFFFFF}                              & \cellcolor[HTML]{FFFFFF}                            & Met.   & 19.43          & 14.71          & 17.44          & 15.10          & 17.65          & 20.09          & 15.58          & 17.21          & 17.15                    \\
\cellcolor[HTML]{FFFFFF}                              & \cellcolor[HTML]{FFFFFF}                            & BLEU   & 23.90          & 20.94          & 23.23          & 20.43          & 22.88          & 19.25          & 20.85          & 21.74          & 21.65                    \\
\cellcolor[HTML]{FFFFFF}                              & \multirow{-3}{*}{\cellcolor[HTML]{FFFFFF}LLaMA-7B}  & Rou.   & 14.01          & 12.14          & 15.01          & 12.61          & 14.90          & 13.36          & 12.60          & 14.49          & 13.64                    \\
\cellcolor[HTML]{FFFFFF}                              & \cellcolor[HTML]{FFFFFF}                            & Met.   & 17.09          & 13.88          & 16.27          & 13.21          & 15.47          & 18.75          & 14.94          & 15.39          & 15.63                    \\
\cellcolor[HTML]{FFFFFF}                              & \cellcolor[HTML]{FFFFFF}                            & BLEU   & 25.97          & 24.80          & 25.39          & 23.04          & 24.32          & 21.13          & 22.53          & 24.45          & 23.95                    \\
\cellcolor[HTML]{FFFFFF}                              & \multirow{-3}{*}{\cellcolor[HTML]{FFFFFF}LLaMA-13B} & Rou.   & 15.10          & 14.04          & 15.52          & 13.04          & 14.70          & 14.33          & 12.77          & 14.53          & 14.25                    \\
\cellcolor[HTML]{FFFFFF}                              & \cellcolor[HTML]{FFFFFF}                            & Met.   & 16.53          & 11.87          & 15.33          & 12.76          & 13.76          & 13.70          & 11.84          & 11.14          & 13.37                    \\
\cellcolor[HTML]{FFFFFF}                              & \cellcolor[HTML]{FFFFFF}                            & BLEU   & 26.16          & 24.96          & 11.01          & 24.20          & 24.78          & 18.97          & 22.49          & 24.43          & 22.13                    \\
\cellcolor[HTML]{FFFFFF}                              & \multirow{-3}{*}{\cellcolor[HTML]{FFFFFF}Gemini}    & Rou.   & 17.62          & 14.34          & 6.30           & 15.13          & 15.97          & 14.46          & 13.90          & 15.50          & 14.15                    \\
\cellcolor[HTML]{FFFFFF}                              & \cellcolor[HTML]{FFFFFF}                            & Met.   & \textbf{21.19} & \textbf{18.00} & \textbf{20.3}  & \textbf{17.05} & \textbf{18.94} & \textbf{21.55} & \textbf{16.32} & \textbf{17.07} & \textbf{18.80(↑78.79\%)} \\
\cellcolor[HTML]{FFFFFF}                              & \cellcolor[HTML]{FFFFFF}                            & BLEU   & \textbf{27.72} & \textbf{26.86} & \textbf{27.25} & \textbf{24.99} & \textbf{27.15} & \textbf{22.68} & \textbf{25.58} & \textbf{26.97} & \textbf{26.15(↑31.42\%)} \\
\multirow{-12}{*}{\cellcolor[HTML]{FFFFFF}LLM}        & \multirow{-3}{*}{\cellcolor[HTML]{FFFFFF}GPT-3.5}   & Rou.   & 19.04          & 17.80          & 19.10          & \textbf{16.70} & 19.04          & 18.31          & 17.15          & \textbf{19.05} & 18.27(↓1.40\%)           \\ \bottomrule
\end{tabular}
}
\vspace{-2em}
\end{table}

\vspace{-0.3cm}

\section{Results and discussion}

\subsection{RQ1: How is the performance of recent LLMs against current CMG approaches?}
\label{RQ1: How is the performance of recent LLMs against current CMG approaches?}

Table \ref{nlpresult} re-evaluates the performance of state-of-the-art CMG approaches of diverse categories against various recent LLMs on MCMDEval+.
When comparing CMG approaches, retrieval-augmented approaches (e.g., RACE, NNGen, and Corec) generally outperform other approaches. The results may be attributed to the fact that similar code \textit{diff}s tend to have similar commit messages, which can supplement effective guidance for these models' generation. 
In particular, NNGen, a purely retrieval-based tool, surpasses most learning-based approaches, highlighting the critical role of retrieving similar examples as guidance in the CMG task. Note that compared to other PLs, the performance of CMG approaches declines more or less in Go, PHP, and Rust. Because the training set of CommitChronicle after preprocessing mentioned in Section \ref{Construction of the high-quality test set} becomes smaller than MCMD, which impacts the model's training and retrieving effectiveness. Additionally, according to Aleksandra et al. \cite{eliseeva2023commit}, CommitChronicle underwent more rigorous data cleaning, such as removing all commits whose authors are present in either the validation or test set from the training set to prevent any overlap even different projects with the same authors. In contrast, this phenomenon is almost nonexistent in LLMs, as these code \textit{diff}s were crawled from VCS in the same manner as MCMD, and the generation by LLMs relies solely on their inherent generalization capabilities, independent of the training set distribution.

Furthermore, when comparing CMG methods with LLMs, LLMs exhibit a substantial advantage in multiple metrics. Because LLMs, such as GPT-3.5, possess a more significant number of parameters and are trained on much more extensive human-written code/text, which enables them to demonstrate more robust performance even without specialized training for specific tasks. However, although RACE has much fewer parameters than LLMs, it surpasses most LLMs in terms of the ROUGE-L metric and performs neck-to-neck with GPT-3.5 on this metric.
One possible explanation is that ROUGE-L emphasizes the longest continuous sequences in whole text sequences. Since RACE is retrieval-based, it can more easily find examples in the corpus that are similar to the target code \textit{diff}. This capability allows RACE to generate high-quality, long-range continuous sequences that maintain a high degree of textual order consistency during commit message generation.
Focusing on the comparison between GPT-3.5 and RACE, we find that GPT-3.5 significantly outperforms RACE, the latest state-of-the-art CMG approach, with an improvement of 78.79\% in terms of METEOR and 31.42\% in terms of BLEU, which completely opposite to the experimental results in the pilot analysis, showing that the original CMG related dataset (e.g., MCMD) indeed distorts the performance evaluation among models and the construction of the cleaned test set is necessary.

In the comparison among LLMs, GPT-3.5 surpasses other LLMs in most evaluation metrics, which may be attributed to its larger number of parameters. 
Regarding the comparison between GPT-3.5 and Gemini, the conclusions of this study are consistent with those of Akter et al. \cite{akter2023depth}, indicating that GPT-3.5 has a relatively clear advantage in tasks related to coding. To be specific, in terms of the METEOR metric, GPT-3.5 leads by an average of 9.62\%$\sim$40.61\% compared to other LLMs. In the BLEU metric, GPT-3.5's average lead ranges from 9.19\% to 20.79\%. Furthermore, in the ROUGE-L metric, GPT-3.5 maintains an average lead of 28.21\%$\sim$33.94\% over its counterparts.

\vspace{-0.2cm}
\begin{boxK}
  \faIcon{pencil-alt}  \textbf{Answer to RQ1:} LLMs have demonstrated impressive performance in the CMG domain from practitioners' perspective, overall surpassing all current CMG approaches, and GPT-3.5 performs the best.
\end{boxK}

\vspace{-0.2cm}

\subsection{RQ2: How accurate is the commit message generated by LLMs?}
\label{RQ2: How accurate is the commit message generated by LLMs?}

\begin{figure*}[!t]
\setlength{\abovecaptionskip}{0cm}
\vspace{-1em}
  \centering
  % \hspace*{-0.65cm}
  \includegraphics[width=\linewidth]{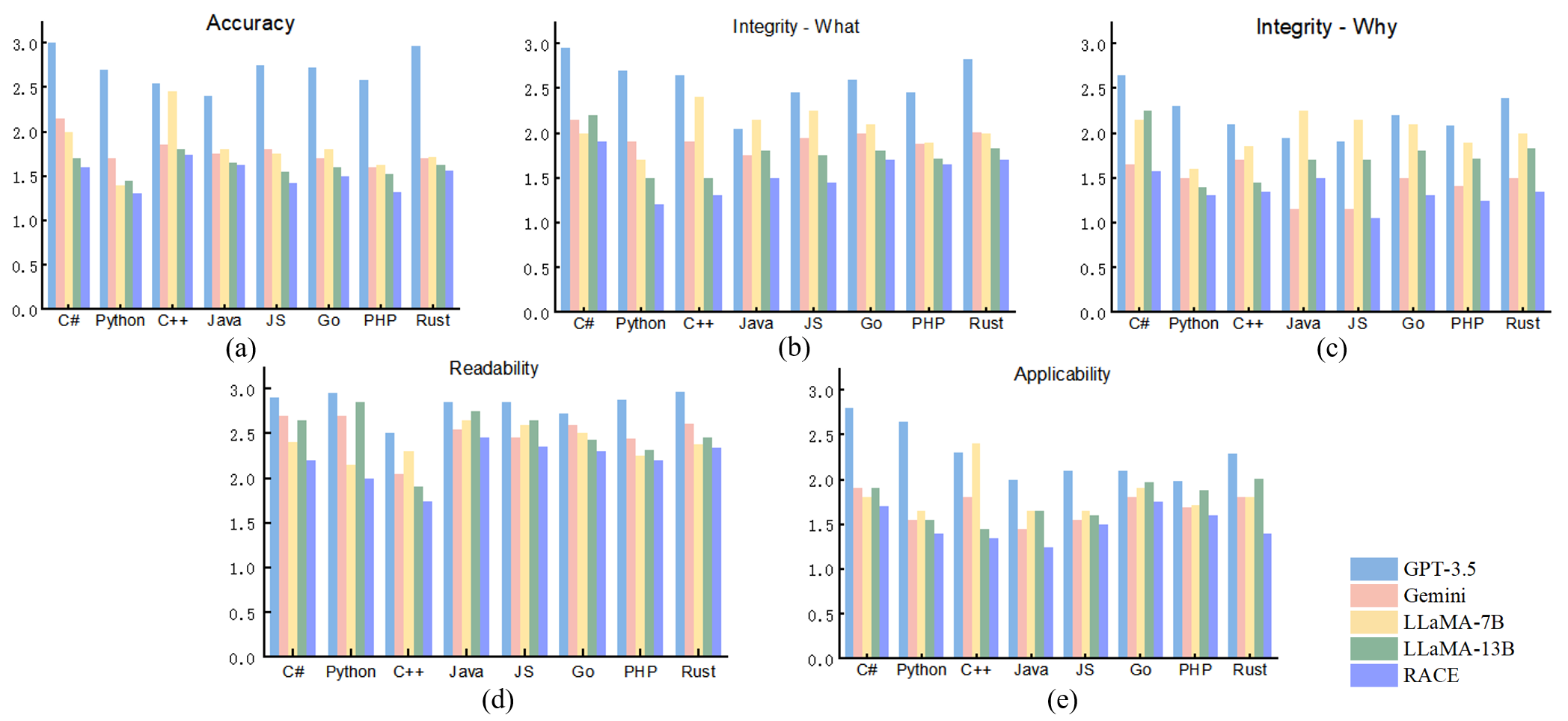}
  \caption{Manual Assessment Results in Terms of Accuracy, Integrity, Readability, and Applicability in Order.}
  \label{metric}
\end{figure*}

Fig. \ref{metric} (a) demonstrates the manual assessment results in terms of Accuracy, where GPT-3.5 still performs the best among different LLMs and receives an average score of 2.71 among evaluators, while the other three LLMs have mean scores of 1.78, 1.82, and 1.61, respectively. RACE receives an average score of 1.51, which is lower than that of all the other LLMs.
Compared with the automated metrics, Accuracy, as a manual metric, measures the overall semantic equivalency instead of rigid literal matching, showing that commit messages generated by LLMs, especially GPT-3.5, perform much better in carrying the main idea of code \textit{diff}s.
In addition, we note that LLaMA-13B outperforms LLaMA-7B overall according to Table \ref{nlpresult}, yet it underperforms LLaMA-7B in terms of the aspect of Accuracy during the manual evaluation. Considering the focus on the semantic equivalency of the Accuracy, we conjecture that LLaMA-7B is superior in generating high-quality commit messages from the perspective of overall semantics, while LLaMA-13B prefers to generate commit messages with higher literal accuracy.

\begin{figure}[!t]
\setlength{\abovecaptionskip}{0cm}
\vspace{-1em}
  \centering
  \includegraphics[width=\linewidth]{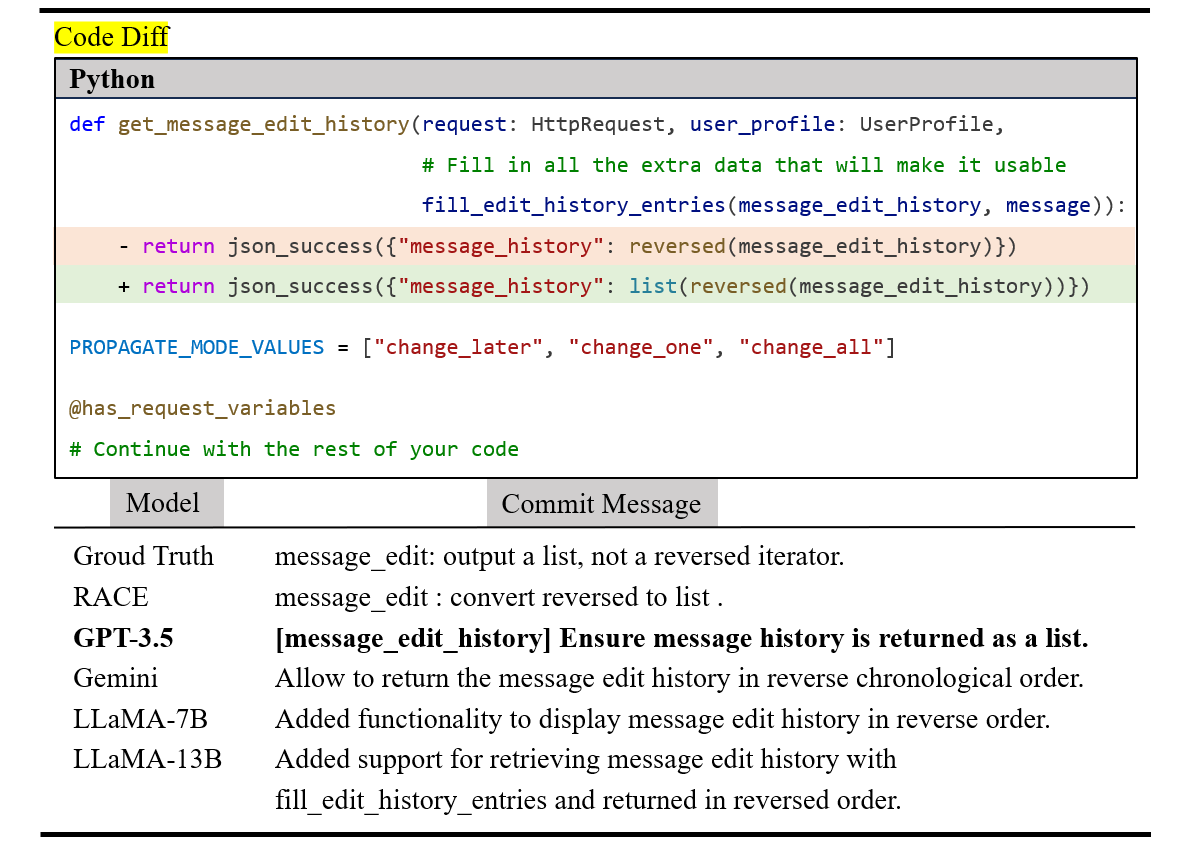}
  \caption{A Python Example of Generated Commit Messages on Accuracy.}
    \label{fig:python}
\vspace{-0.4cm}
\end{figure}

Taking a Python code \textit{diff} as an example, we list different models' generated commit messages in Fig. \ref{fig:python}. It can be observed that among the commit messages generated by RACE and the four LLMs, the commit message generated by GPT-3.5 directly and clearly emphasizes the core content of the code change, i.e., ensuring that the message history is returned in list form. It conveys the essence of the code change without introducing unnecessary information or incorrect statements, aligning more closely with the original code change situation. Looking at the commit messages generated by the other three LLMs for this code change, we find that these commit messages only mention ``return in reverse order'' and do not explicitly state the change key of ``return in the form of a list.'' In contrast, although the commit message generated by RACE mentions the keyword ``list'', it fails to convey the correct intent of the code \textit{diff}, and the description of the object being modified is inaccurate.

\begin{boxK}
 \faIcon{pencil-alt} \textbf{Answer to RQ2:} Manual assessment has demonstrated that GPT-3.5 exhibits the best performance among all LLMs and RACE in terms of Accuracy in the CMG task, showcasing its exceptional ability to capture semantic equivalency and accurately reflect code changes. 
\end{boxK}
\vspace{-0.2cm}

\vspace{-0.3cm}

\subsection{RQ3: How integral is the commit message generated by LLMs?}
\label{RQ3: How integral is the commit message generated by LLMs?}
\begin{figure}[!t]
\vspace{-1em}
  \centering
  \setlength{\abovecaptionskip}{0cm}
  \includegraphics[width=\linewidth]{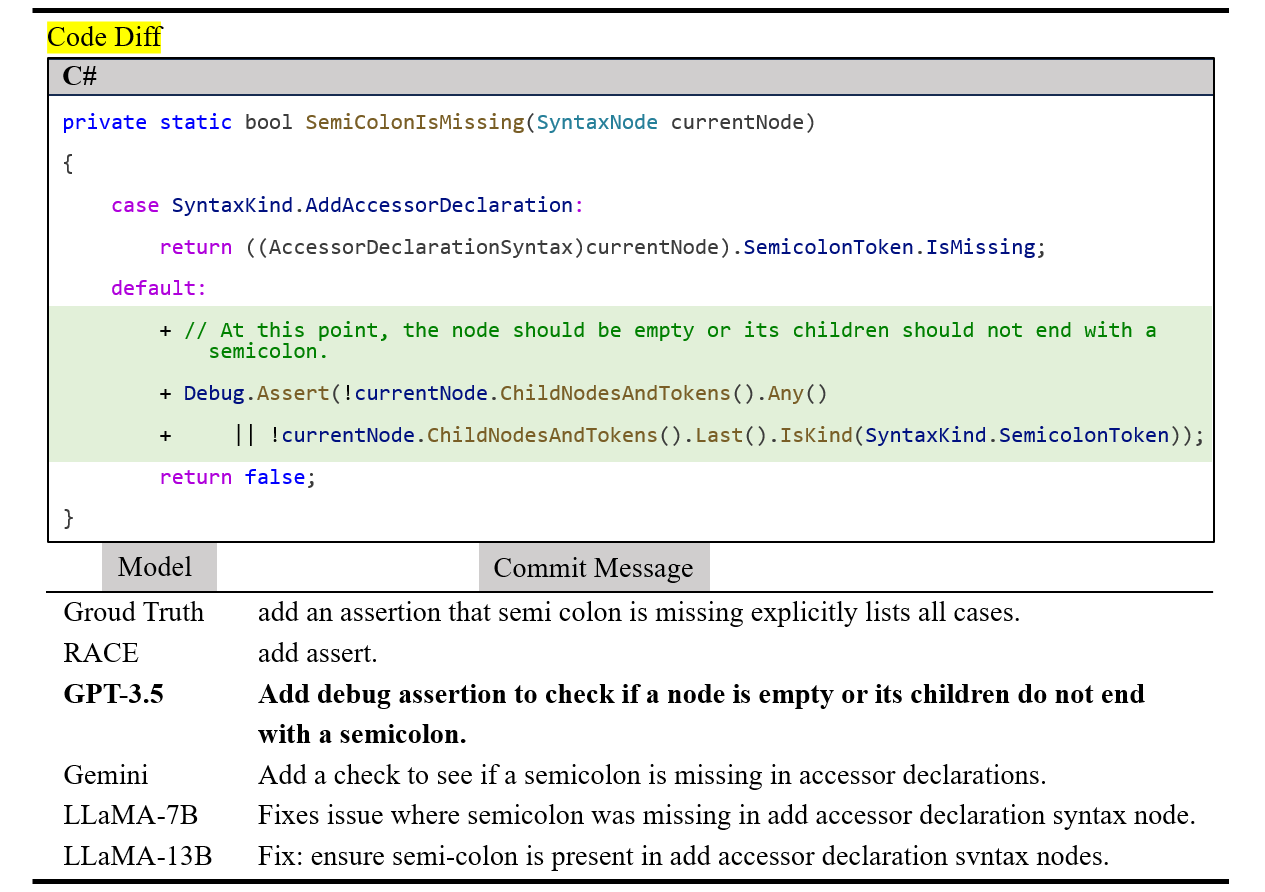}
  \caption{An C\# Example of Generated Commit Messages on Integrity/Applicability.}
    \label{C}
\vspace{-1.5em}
\end{figure}

Fig. \ref{metric} (b) (c) demonstrates the Integrity concerning ``What'' and ``Why'' information of the commit messages generated by different models.
Firstly, we focus on evaluating commit messages in the ``What'' aspect. As can be seen, GPT-3.5 excels with a mean score of 2.59, while the other three LLMs have mean scores of 1.94, 2.06, and 1.76, respectively. Besides, RACE has the lowest average score at 1.55. This indicates that commit messages generated by GPT-3.5 normally convey in detail the actions code changes have taken. Fig. \ref{C} demonstrates an example of a code \textit{diff} in C\#, and it is observed that among the commit messages generated by the five models, GPT-3.5-generated commit messages provides a clear description of the code changes (i.e., ``What'' information), namely ``add debug assertion''. Nonetheless, the corresponding contents in commit messages generated by other models are relatively vague, such as Gemini and LLaMA-7/13B, or simplified, such as RACE. 

Secondly, we examine the evaluation results of model-generated commit messages from the ``Why'' aspect. Apparently, GPT-3.5 also performs exceptionally well, with a mean score of 2.20, while Gemini, LLaMA-7/13B, and RACE receive mean scores of 1.45, 2.00, 1.73, and 1.33, respectively. This indicates that commit messages generated by GPT-3.5 more clearly explain the reasons and motivations behind the code changes. Using the aforementioned CMG example in C\# for illustration, among the commit messages generated by the five models, GPT-3.5-generated commit messages contain more explicit ``Why'' information, i.e.,  ``check if a node is empty or its children do not end with a semicolon,'' while the ``Why'' information in the commit messages generated by the other models are relatively vague (e.g., LLaMA-7/13B), incorrect (e.g., Gemini) or even absent (e.g., RACE), as shown in Fig. \ref{C}.

It is noteworthy that all LLMs examined generally score higher in terms of the ``What'' aspect than the ``Why'' aspect when generating commit messages.
This phenomenon may stem from the limitations of the code snippets themselves. Actually, in development practice, code snippets rely on each other closely and have their own specific functionalities, which can only be reflected when they are placed in the whole repository. Nonetheless, code \textit{diff} samples in the MCMDEval+ dataset only comprise fragments. In this case, although LLMs can easily understand the specific content of code changes, they can hardly capture the motivation behind code \textit{diff}s owing to the lack of adequate context information of the whole repository.

\begin{boxK}
  \faIcon{pencil-alt} \textbf{Answer to RQ3:} GPT-3.5 excels in generating integral commit messages with both ``What'' and ``Why'' information, significantly outperforming other LLMs. All models perform relatively weaker in expressing ``Why'' information in commit messages owing to the limited code context.
\end{boxK}
\vspace{-0.2cm}

\vspace{-0.3cm}

\subsection{RQ4: How readable is the commit message generated by LLMs?}

\begin{figure}[!h]
\setlength{\abovecaptionskip}{0cm}
\vspace{-1em}
  \centering
  \includegraphics[width=\linewidth]{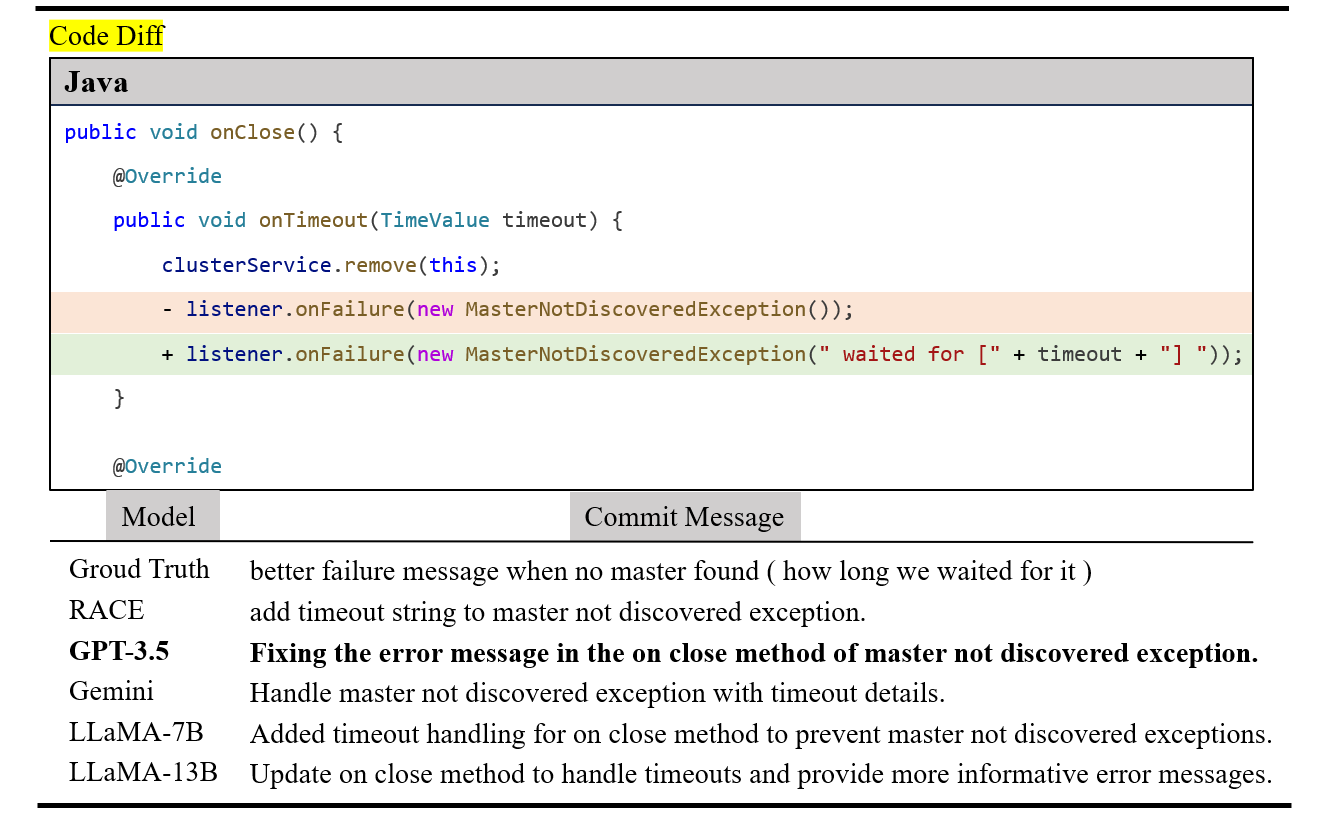}
  \caption{A Java Example of Generated Commit Messages on Readability.}
    \label{Java}
\vspace{-1em}
\end{figure}

Fig. \ref{metric} (d) demonstrates the Readability assessment for diverse models, where GPT-3.5, Gemini, LLaMA-7B, LLaMA-13B, and RACE have mean scores of 2.83, 2.51, 2.40, 2.50, and 2.20 respectively, where GPT-3.5 has a slight advantage, while the other three LLMs perform comparably and RACE has a slightly lower performance.
Simultaneously, we observe that, among all test dimensions, Readability scores are generally higher and distributed relatively evenly compared to scores in other aspects, indicating the powerful capabilities of recent LLMs and RACE in fluent expression and good grammar when generating commit messages. Taking a Java code \textit{diff} as an example (as shown in Fig.~\ref{Java}), all model-generated commit messages clearly expressed their own understanding of the given code \textit{diff} in general.
As one evaluator commented, ``Overall, each commit message in this example is fluent and easy to understand without any obvious grammar error, demonstrating a good readability. However, I particularly appreciate the commit messages generated by GPT-3.5, which succinctly and fluently describe the specific issues being addressed.".

\begin{boxK}
 \faIcon{pencil-alt} \textbf{Answer to RQ4:} Manual assessments have found that both LLMs and RACE perform well in terms of Readability. Among them, GPT-3.5 has a slight edge, reflecting its natural and fluent conveyance of code change information. 
\end{boxK}

\vspace{-0.4cm}

\subsection{RQ5: How applicable is the commit message generated by LLMs?}
\label{RQ5: How applicable is the commit message generated by LLMs?}

Fig. \ref{metric} (e) illustrates the manual assessment results concerning the Applicability of different models by PLs, indicating that GPT-3.5 excels in Applicability compared to Gemini, LLaMA-7B, LLaMA-13B, and RACE. Among these models, the mean scores are 2.28 for GPT-3.5, 1.69 for Gemini, 1.82 for LLaMA-7B, 1.75 for LLaMA-13B, and 1.49 for RACE.
For example, when using the aforementioned Java CMG example (as shown in Fig.~\ref{C}), the commit message generated by GPT-3.5 outperforms those generated by the other four models across various aspects. It accurately and fluently conveys the core of the code change while clearly incorporating the essential ``What" and ``Why" elements. Even though it is not literally consistent with the ground truth, they manifest the same meaning. In contrast, other models lack either ``What'' or ``Why'' information, as mentioned in Section \ref{RQ3: How integral is the commit message generated by LLMs?}. Besides, LLaMA-7B and LLaMA-13B are relatively less readable while Gemini does not express clearly the modification target, namely ``assertion''. 
These above omissions may reduce developers' willingness to adopt these commit messages, thereby negatively impacting their applicability.

\begin{boxK}
 \faIcon{pencil-alt} \textbf{Answer to RQ5:} Compared to other models, GPT-3.5 demonstrates higher Applicability in generating commit messages. Its concise and accurate descriptions effectively and fluently convey the details of code changes, thereby gaining higher acceptance among evaluators.
\end{boxK}

\vspace{-0.3cm}

\subsection{Discussion}
\begin{figure}[!h]
\setlength{\abovecaptionskip}{0cm}
\vspace{-1em}
  \centering
  \includegraphics[width=\linewidth]{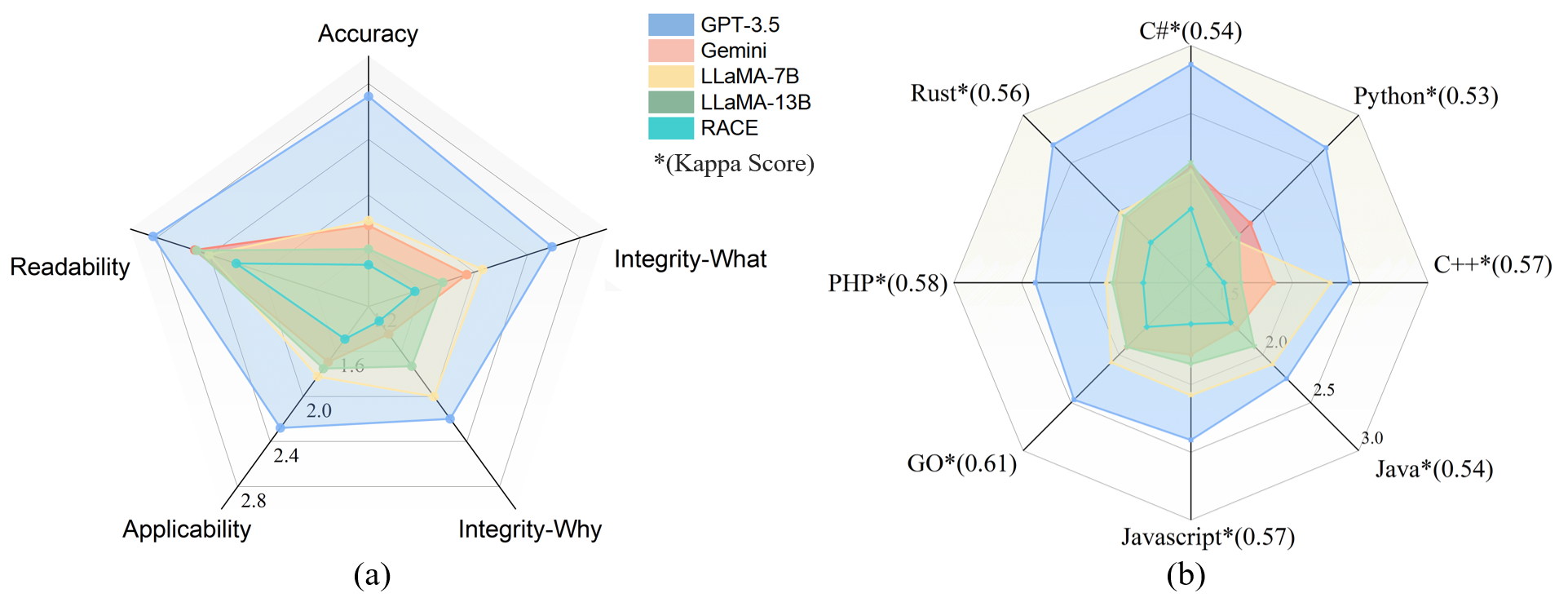}
    \caption{An Overview of Various Models' Performance in Different Aspects of Manual Assessment and on Different PLs}
    \label{RA}
\vspace{-0.8em}
\end{figure}

\textbf{Performance of different models on manual metrics.} Fig. \ref{RA} (a) presents the average performance of various LLMs and RACE among different aspects of the manual assessment. It can be observed that, compared to RACE, LLMs consistently outperform across all aspects. Particularly, GPT-3.5 performs best overall, while the rest of the LLMs manifest diverse advantages. For example, LLaMA-7B demonstrates relative superiority in terms of Integrity-Why, and most LLMs can generate quite readable commit messages.

\textbf{Performance of different models on diverse PLs.} Fig. \ref{RA} (b) illustrates the average scores of different models on various PLs based on manual assessment. Additionally, the Cohen's kappa coefficient between the two experts for each PL in Section \ref{Evaluation Methodology} is calculated and is displayed in the figure. The Cohen’s kappa coefficient for each PL exceeds 0.5, indicating a ``substantial agreement'' result, indicating the effectiveness of the manual evaluation results of Section \ref{RQ2: How accurate is the commit message generated by LLMs?}-\ref{RQ5: How applicable is the commit message generated by LLMs?}. 
Consistent with previous findings, GPT-3.5 still achieves the best performance across all PLs. Among the four selected LLMs, three (i.e., GPT-3.5, Gemini, and LLaMA-13B) perform best in C\#, and LLaMA-7B excels in C++. This suggests that different LLMs have distinct advantages for different PLs. Although there are some discrepancies with the results from automated evaluations, the relative performance differences between models remain largely consistent.

\section{ERICommiter}
In the previous six sections, we conduct an empirical study on the use of LLMs for the automatic generation of commit messages. Our evaluation, through both automated and manual analysis, highlighted the superiority of LLMs in producing accurate, relevant, and informative commit messages compared with state-of-the-art CMG approaches of diverse categories. In this section, considering the prominent efficacy of incorporating retrieved samples for references in CMG approaches (as mentioned in Section \ref{RQ1: How is the performance of recent LLMs against current CMG approaches?}) and the extensive applicability of ICL for LLMs \cite{dong2022survey,lu2024grace,yang2023auto}, we carry an idea of retrieving similar samples towards the target code \textit{diff} to construct ICL examples, thereby guiding LLMs' generation and further improving their performance in the CMG task. However, the volume of the database for retrieval is exceptionally huge and full of noise. For example, each PL in the MCMD dataset contains over 200,000 samples, but not all samples are good candidates for retrieval, while the retrieval database in practice can be even larger. To accelerate the retrieval efficiency and make this approach practical, we propose a two-step filtering method to reduce the database volume before retrieving and ensure the CMG performance remains almost the same concurrently. To summarize, we introduce the Efficient Retrieval-based In-context Learning framework, named ERICommiter, suitable for different LLMs in the CMG field. Here's how it works: (1) We employ a two-step filtering process to eliminate less informative and low-quality samples, constructing the reduced retrieval database. (2) Based on the reduced database, we retrieve similar samples to build ICL examples. (3) LLMs are guided to generate more accurate and informative commit messages via ICL. 

\vspace{-0.2cm}

\subsection{Methodology}
\subsubsection{Two-Step Filtering on the Retrieval Database}

This stage includes two steps for sample filtering from the retrieval database as presented below, namely length-based filtering and semantic-based filtering.

\begin{figure}[!t]
\setlength{\abovecaptionskip}{0cm}
\vspace{-1em}
  \centering
  \includegraphics[width=\linewidth]{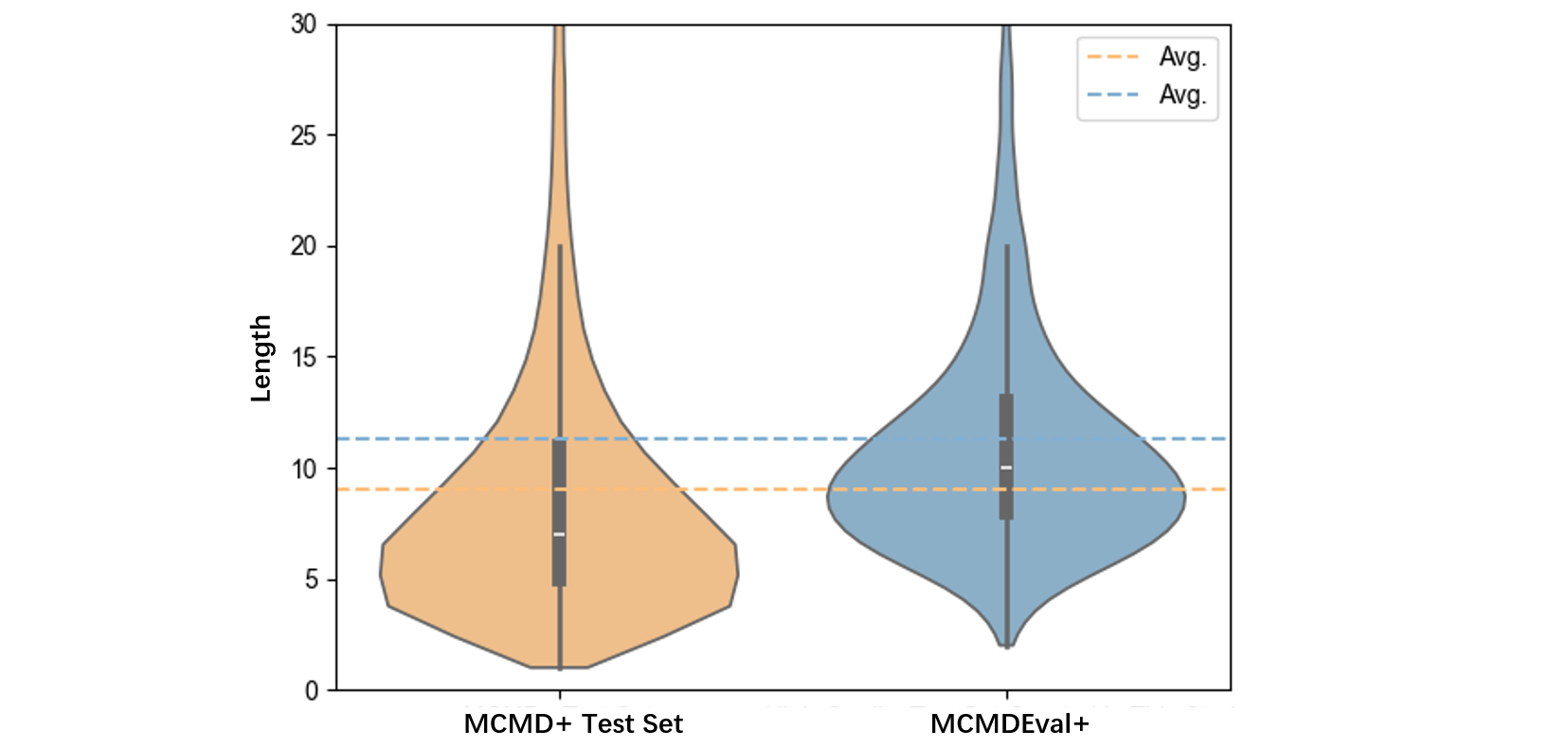}
  \caption{Comparison of Length in Commit Messages.}
    \label{token}
\vspace{-1.5em}
\end{figure}

\textbf{Length-based Filtering:}
An intuitive hypothesis suggests that high-quality commit messages are relatively longer (i.e., containing more tokens) because they typically include both ``What" and ``Why" information about code \textit{diff}s, such as the specific functionalities introduced or modified, the implementations used, and the underlying motivations for the changes. To substantiate this hypothesis, we plot violin diagrams to compare the distribution of token counts in the test set of MCMD+ and our cleaned high-quality test set MCMDEval+, as shown in Fig. \ref{token}. The distribution discrepancies in token counts between them are obvious; the cleaned high-quality test set carries more tokens for each commit message on average, showing that length-based filtering on the retrieval database is reasonable. In this experiment, we adopt the average length of the cleaned high-quality test set as the threshold to filter samples in the MCMD+'s training set (i.e., retrieval database).

\textbf{Semantic-based Filtering:}
Secondly, to further filter out high-quality commit messages, we again utilize the Bi-LSTM constructed in Section \ref{Construction of the high-quality test set} to capture and understand the critical elements of ``What'' and ``Why'' in commit messages, thereby effectively filtering out high-quality commit messages semantically.

These two filtering steps are applied to the original training set. The first step, based on the length of commit messages, eliminates overly short commit messages, and the second step uses the Bi-LSTM model to further filter out high-quality samples from a semantic perspective. 
Through these two filtering steps, we obtain a much smaller but relatively high-quality training set as the reduced retrieval database. 

\subsubsection{Retrieval Method} 
\label{Retrieval Method}
In this study, we investigate two commonly used retrieval methods in academia, namely lexical-based retrieval and semantic-based retrieval. These methods are used to select commit messages similar to the test samples from the training set. 

\textbf{Lexical-based Retrieval:} uses the BM25 algorithm \cite{robertson2009probabilistic}, a widely-used information retrieval function based on the bag-of-words model \cite{zhang2010understanding}, evaluating the relevance between documents and queries by considering Term Frequency (TF) and Inverse Document Frequency (IDF). This method excels at keyword matching and is suitable for quickly retrieving documents related to specific query conditions (such as specific PLs or keywords) in extensive text collections.

\textbf{Semantic-based Retrieval:} exerts a pre-trained code model, namely CodeReviewer \cite{li2022codereviewer}, a Transformer-based encoder-decoder model designed with four pre-training tasks for the code review process. CodeReviewer takes code \textit{diff}s as input and splits them into token sequences using the RoBERTa tokenizer. 
To enhance comprehension of the code \textit{diff} format, we replace the special markers ``-" and ``+" in the code \textit{diff} file, which represent line deletion and insertion, with the special tokens [DEL] and [ADD], respectively. Additionally, lines that remain unchanged, indicated by the absence of any marker, are denoted by the token [KEEP].
The output of the model includes token representations from the encoder and the generated token sequence from the decoder. In this study, we only use the pre-trained encoder and the last layer representation of the special token [CLS] to represent the vectorized code \textit{diff}. First, we calculate the cosine angle between two [CLS] token representations to measure the similarity between code changes. This approach offers an effective method for assessing code modifications. Second, to reduce the computational overhead, we limit the number of retrieved examples to one in our experiments.

\vspace{-0.3cm}

\subsection{Experimental Setting of ERICommiter}
To investigate the effectiveness and efficiency of ERICommiter, we further propose RQ6$\sim$RQ8 and their evaluation procedures below.

\noindent\textbf{RQ6: How does ERICommiter perform against individual LLMs?}
This RQ aims to examine the effectiveness of ERICommiter compared with individual LLMs tested. To make a fair comparison, we retrieve the most similar sample from the reduced retrieval database for ERICommiter while selecting one fixed example for individual LLMs. Due to computational overhead and resource constraints, we include two PLs (i.e., Java and Python) and two LLMs (i.e., GPT-3.5 and Gemini) for experiments, both semantic and lexical-based retrieval methods are assessed. 

\noindent\textbf{RQ7: How efficient is ERICommiter?}
This RQ aims to evaluate the impact of the two-step filtering mechanism on the efficiency and performance of ERICommiter. Following the settings of RQ6, we conduct an ablation study by systematically disabling components of the filtering mechanism. Specifically, we compare the performance of ERICommiter with two modified versions: one with only the second step of filtering removed, namely ERICommiter w/o Step 2, and another with both steps of filtering removed, namely ERICommiter w/o Step 1\&2. These ablation experiments provide a deeper understanding of how each filtering step contributes to the overall efficiency and effectiveness of ERICommiter.

\noindent\textbf{RQ8: How does the number of retrieved examples influence the performance of ERICommiter?}
Based on the experimental setting in RQ6, we further alter the number of retrieved examples $N$=$\{1, 3, 5, 10\}$ to evaluate its influence on ERICommiter. Notably, due to the input token limit of GPT-3.5-turbo (4,096 tokens at most), we could only feed one example for it. However, to more comprehensively understand the impact of the number of examples on LLMs' generation performance, we introduce the version of GPT-3.5-16k as an additional experimental subject, which can accommodate 16,385 tokens for input. Although the GPT-3.5-16k may not perform as well as the GPT-3.5-turbo overall, the two LLMs are consistent in basic principles. 
Therefore, when evaluating the influence of the example number on LLM with ICL, we can replace GPT-3.5-turbo with GPT-3.5-16k to observe the performance variation for GPT-3.5 series LLMs.

\begin{table}[h]
\setlength{\abovecaptionskip}{0cm}
\vspace{-1em}
\footnotesize
\centering
\caption{Performance Comparison of ERICommiter Against Various Underlying LLMs}
\label{result}
\renewcommand{\arraystretch}{1.25}
\resizebox{\linewidth}{!}{
\begin{tabular}{lccc|ccc}
\toprule
\multicolumn{1}{c}{}                        & \multicolumn{3}{c}{Java}                                       & \multicolumn{3}{c}{Python}                                      \\ \cline{2-7} 
\multicolumn{1}{c}{\multirow{-2}{*}{Model}} & Met.                         & BLEU                     & Rou.              & Met.                         & BLEU                     & Rou.                            \\ \hline
\multicolumn{7}{c}{Lexicial-based Retrieval}\\\hline
GPT-3.5 &20.40& 28.27& 18.50& 17.02 &28.96& 18.33\\
ERICommiter(GPT-3.5)                         & \textbf{23.54}                        & \textbf{30.40}                               & \textbf{21.98}                                & \textbf{19.60}                               & \textbf{29.53}                             & \textbf{19.91}                              \\

Gemini&5.96& 11.24 &6.66& 3.65& 7.59 &5.22

\\

ERICommiter(Gemini) & \textbf{16.00} & \textbf{23.76} & \textbf{17.18} & \textbf{12.06} & \textbf{22.72} & \textbf{14.49} \\

\hline
\multicolumn{7}{c}{Semantic-based Retrieval}\\\hline
GPT-3.5 &20.40& 28.27& 18.50& 17.02 &28.96& 18.33\\
ERICommiter(GPT-3.5)   & \textbf{21.81} & \textbf{29.89} & \textbf{20.13} & \textbf{19.03} & \textbf{29.05} & \textbf{19.55} \\
Gemini&5.96& 11.24 &6.66& 3.65& 7.59 &5.22
\\
ERICommiter(Gemini)  & \textbf{15.54} & \textbf{24.19} & \textbf{17.44} & \textbf{11.20} & \textbf{20.86} & \textbf{13.17} \\ 
\bottomrule
\end{tabular}
}
\vspace{-2.5em}
\end{table}

\subsection{Experimental Results of ERICommiter}

\textbf{RQ6: How does ERICommiter perform against individual LLMs?}
Table \ref{result} showcases the experimental results between ERICommiter and its underlying LLMs with lexical/semantic-based retrieval methods, respectively. Apparently, ERICommiter improves LLMs' CMG performance across different PLs consistently. Utilizing a lexical-based retrieval approach, GPT-3.5 achieves average enhancements of 15.26\%, 4.75\%, and 13.72\% in the METEOR, BLEU, and ROUGE-L, respectively. Besides, Gemini exhibits substantial improvements of 199.43\%, 155.36\%, and 167.77\% in terms of each evaluation metric in order. As for the setting of semantic-based retrieval, GPT-3.5 demonstrates average gains of 9.36\%, 3.02\%, and 7.73\% in terms of METEOR, BLEU, and ROUGE-L, respectively. Meanwhile, Gemini shows significant enhancements of 183.79\%, 145.02\%, and 157.08\% in terms of each metric in order. In summary, similar examples significantly improve the effectiveness of ICL for CMG with LLMs. 

\begin{figure}[h]
\setlength{\abovecaptionskip}{0cm}
\vspace{-1em}
  \centering
  % \hspace*{-0.65cm}
  \includegraphics[width=\linewidth]{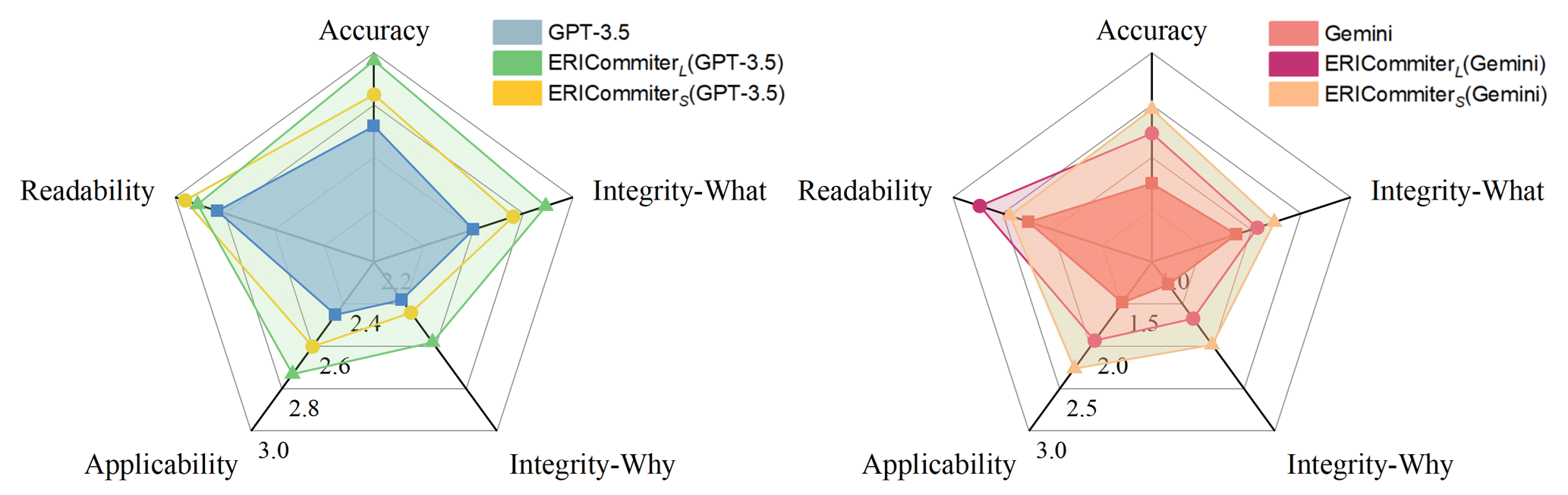}
  \caption{An Overview of ERICommiterr's Performance Among Different Aspects of Manual Assessment }
  \label{eriresult}
\vspace{-1.5em}
\end{figure}

To better illustrate ERICommiter’s performance compared to individual LLMs that only include fixed examples in the prompt, we further evaluate the performance of ERICommiter following the manual evaluation procedure described in Section \ref{Evaluation Methodology}. We present radar charts in Fig. \ref{eriresult}, showcasing the enhancement of commit message quality across five key aspects after employing ERICommiter on GPT-3.5 and Gemini with lexical/semantic-based retrieval methods. For example, ERICommiter$_L$(GPT-3.5) denotes the results of ERICommiter with GPT-3.5 leveraging the lexical retrieval method. To be specific, both GPT-3.5 and Gemini demonstrate significant improvements with ERICommiter in terms of all aspects, especially Accuracy and Applicability, when generating commit messages. 
Judging from all the above, ERICommiter substantially enhances LLMs' CMG performance by retrieving high-quality similar examples from real-world projects for ICL. Because similar code \textit{diff}s normally carry referable commit messages as examples, whereas a fixed ICL example may be irrelevant and cannot guide LLMs' inference in the CMG task. 

\begin{boxK}
 \faIcon{pencil-alt} \textbf{Answer to RQ6:} Our proposed framework, namely ERICommiter, consistently and substantially improves its underlying LLMs' performance when generating commit messages for code \textit{diff}s in diverse PLs and aspects. 
\end{boxK}

\begin{table}[h]
\vspace{-2em}
\setlength{\abovecaptionskip}{0cm}
\footnotesize
\centering
\caption{Performance Evaluation on the Two-step Filtering Module}
\label{result2}
\renewcommand{\arraystretch}{1.35}
\resizebox{\linewidth}{!}{
\begin{threeparttable}
\begin{tabular}{lccc|ccc|c}
\toprule
\multicolumn{1}{c}{}                        & \multicolumn{3}{c}{Java}                                       & \multicolumn{3}{c|}{Python}  &AVG                          \\ \cline{2-7} 
\multicolumn{1}{c}{\multirow{-2}{*}{Model}} & Met.                         & BLEU                     & Rou.              & Met.                         & BLEU                     & Rou.   &Time(s)                          \\ \hline
\multicolumn{8}{c}{Lexicial-based Retrieval}\\\hline
ERICommiter(GPT-3.5)                     & 23.54                          & 30.40                           & 21.98                           & \textbf{19.60}                      & 29.53                           & \textbf{19.91}     & \textbf{35.99}                       \\

ERICommiter w/o Step 2(GPT-3.5)                       & 23.83                       & 31.75                           & 22.55                           & 18.61                     & 30.20                          & 19.46    & 196.73                      \\

ERICommiter w/o Step 1\&2(GPT-3.5)      & \textbf{24.35} & \textbf{32.11} & \textbf{23.73} & 18.80 & \textbf{30.21} & 19.73  & 594.22               \\
ERICommiter(Gemini)               & 16.00                          & 23.76                           & 17.18                           & 12.06                          & 22.72                           & 14.49        & \textbf{35.99}                 \\
ERICommiter w/o Step 2(Gemini)    &     \textbf{20.21} & \textbf{25.45} & 17.19 & \textbf{14.49} & 20.91 & 14.64
  & 196.73              \\
ERICommiter w/o Step 1\&2(Gemini)    &    16.52 & 25.31 & \textbf{17.34} & 11.72 & \textbf{24.23} & \textbf{16.50} 
  & 594.22                  \\

\hline
\multicolumn{8}{c}{Semantic-based Retrieval}\\\hline
ERICommiter(GPT-3.5)                  & 21.81                          & \textbf{30.89}                           & 20.13                           & \textbf{19.03}                          & 29.05                           & 19.55            & \textbf{0.81}                \\
ERICommiter w/o Step 2(GPT-3.5)     &         22.07 & 30.57 & 21.48 & 17.33 & 29.56 & 19.07  & 3.96                          \\

ERICommiter w/o Step 1\&2(GPT-3.5)     &         \textbf{22.24} & 30.86 & \textbf{22.30} & 18.22 & \textbf{30.06} & \textbf{20.06}  & 11.31                          \\
ERICommiter(Gemini)                & 15.54                         & 24.19                           & 17.44                           & 11.20                          & 20.86                           & 13.17         & \textbf{0.81}                   \\

ERICommiter w/o Step 2(Gemini)  &     \textbf{18.81} & \textbf{24.43} & 17.41 & \textbf{13.41} & \textbf{21.02} & \textbf{13.94} &3.96    \\ 

ERICommiter w/o Step 1\&2(Gemini)  &     14.90 & 23.85 & \textbf{18.48} & 10.01 & 20.79 & 11.31 &11.31     \\ 
\bottomrule
\end{tabular}

\end{threeparttable}
}
\vspace{-1em}
\end{table}

\textbf{RQ7: How efficient is ERICommiter?}
Table \ref{result2} illustrates the efficacy and efficiency comparison between ERICommiter and its ablated versions, where one or both steps of the two-step filtering module have been removed. The results indicate that ERICommiter reduces the retrieval time to 18.29\% of the Step 2 ablated version and 6.06\% of the fully ablated version for lexical-based retrieval, while to 20.45\% of the Step 2 ablated version and 7.16\% of the fully ablated version for semantic-based retrieval, substantially improving the operating efficiency of the framework. Meanwhile, ERICommiter still maintains high performance, being neck-to-neck with the ablated versions. Nevertheless, there are still some cases where ERICommiter slightly underperforms ablated versions, as the two-step filtering process inevitably excludes some more similar and useful examples from the retrieval database. Considering the significantly reduced time overhead that makes ERICommiter a practical framework, in reality, the two-step filtering process is still an invaluable and necessary module.

Besides, the lexical-based retrieval method costs much more time than the semantic-based method. The reason is that the foundation of BM25 is the TF-IDF algorithm as mentioned in \ref{Retrieval Method}. When retrieving, BM25 parses the query into its constituent morphemes $Q=\{q_i\}$. For each candidate $d$ in the retrieval database, it calculates the relevance score of each morpheme $q_i$ in the query with $d$ and sums them up to obtain the overall relevance score. In this way, BM25 can rank all the overall relevance scores in descending order to find the most relevant $d$ as the retrieval result \cite{robertson2009probabilistic}. Thus, BM25 cannot construct a vector database in advance and needs to re-compute all procedures for every retrieval candidate, resulting in slow retrieval speeds. In contrast, semantic retrieval only needs to perform the operation of converting the code \textit{diff} into a fixed-dimensional vector once and then use cosine similarity to measure the relevance, leading to relatively low computational overhead.

\begin{boxK}
 \faIcon{pencil-alt} \textbf{Answer to RQ7:} ERICommiter substantially reduces the retrieval time cost compared with two ablated versions and carries almost the same performance, showing the high practicality of ERICommiter in real software development and maintenance.  
\end{boxK}

\begin{figure*}[!t]
\setlength{\abovecaptionskip}{0cm}
\vspace{-1em}
  \centering
  % \hspace*{-0.65cm}
  \includegraphics[scale=0.15]{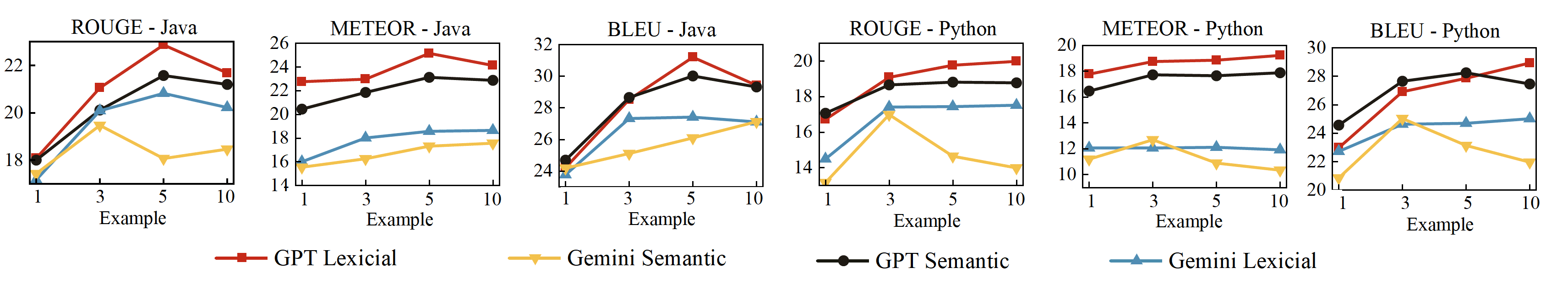}
  \caption{The Influence of the Number of Retrieved Examples on ERICommiter}
  \label{expresult}
% \vspace{-0.9em}
\end{figure*}

\textbf{RQ8: How does the number of retrieved examples influence the performance of ERICommiter?}
Fig. \ref{expresult} demonstrates the results of ERICommiter provided with various retrieved examples. As can be seen, with an increasing number of provided examples, the performance of the LLMs generally shows an upward trend in most situations. As more examples are available, they offer richer contextual information, aiding the model in accurately understanding and generating commit messages relevant to the target code \textit{diff}.
However, this trend does not apply in all cases. In some LLMs, when the number increases excessively, their performance decreases. A potential explanation is more examples may include more noises, as samples that are similar to the target code \textit{diff} in the retrieval database are limited. Therefore, selecting an appropriate number of high-quality examples is crucial for optimizing model performance.

\begin{boxK}
 \faIcon{pencil-alt} \textbf{Answer to RQ8:} The performance of ERICommiter improves with more retrieved examples, but excessive examples can introduce noise and bring about performance decline, demonstrating that retrieving an optimal number of examples is crucial.
\end{boxK}

\vspace{-0.5cm}

\subsection{Case Study}

To thoroughly evaluate the practicality of ERICommiter, we conduct a case study on a university-level open-source project, namely SDUOJ, involving various programming-related teaching supports, such as subjective questions, time-limited exams, and online program judging \cite{SDUOJOnl17:online, SDUOJsdu78:online}. 
This project is maintained by an 11 member-team (i.e., SDUOJ team), all of whom have extensive experience in Java or Python development and place a high value on the quality
of commit messages during their software development process. To support this case study, we designed a two-month experiment in which the SDUOJ team is required to use ERICommiter to generate their commit messages. To do this, we developed ERICommiter as a web application using the Flask framework and deployed it on our private server, which not only provides a user interface but also integrates a feedback mechanism to allow us to track whether the SDUOJ team members accept the commit messages generated by ERICommiter and their feedback. The website code is available at \cite{Pengyu0324:online}.

Specifically, before the experiment commenced, we provided a brief training session to the SDUOJ team members, introducing them to the usage of ERICommiter and the website's operational workflow. During the experiment, each member is required to use ERICommiter to generate commit messages for code submissions through the provided website. If they accept the generated message, they can submit it directly; otherwise, they are required to fill out a short form explaining the reasons for their rejection.

\begin{figure}[!h]
\setlength{\abovecaptionskip}{0cm}
\vspace{-1em}
  \centering
\includegraphics[width=\linewidth]{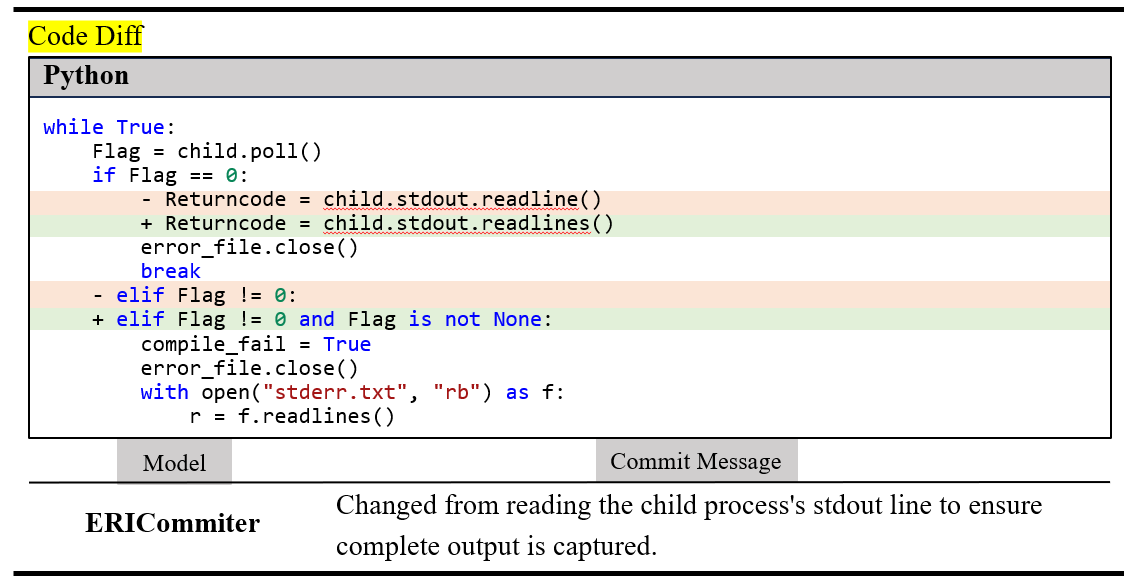}
  \caption{An Example of a Rejected Commit Message Generated by ERICommiter.}
  \label{failure}
\vspace{-0.2em}
\end{figure}

At the end of the experiment, we collected 364 data points, with 98.1\% of the commit messages generated by ERICommiter being accepted. Additionally, we design a survey using a 3-point Likert scale to gather the SDUOJ team members' feedback on ERICommiter. The overall approval rating of the commit messages generated by ERICommiter reached 2.82, and the agreement that these commit messages improve their work efficiency reached 2.91. However, some team members also identify areas where ERICommiter could be improved, primarily noting that the generated commit messages are sometimes not specific enough or focus only on certain parts of the code \textit{diff} while overlooking others. For example, Fig. \ref{failure} shows a rejected commit message and its corresponding code \textit{diff}. As observed, the ERICommiter-generated commit message only addresses part of the code changes that excludes the illustration of a condition update (i.e., from ``elif Flag != 0'' to ``elif Flag != 0 and Flag is not None'').

Through this case study, we have demonstrated ERICommiter's potential for application in a real-world development environment. Although there are areas needing improvement, overall, ERICommiter significantly enhances the quality of commit messages and improves the development efficiency.

\vspace{-0.3cm}

\section{Threats to Validity}

\textbf{Internal Validity:} One threat to the internal validity lies in the potential of data leakage that the cleaned dataset may have a certain overlap with the training samples of LLMs examined. As samples in MCMDEval+ were crawled from Github, it is inevitable that LLMs might have seen some project code during their pre-training stage. However, those samples are all in formats of code \textit{diff}s owing to the characteristic of the CMG task, which carries a totally different input format against the pre-training samples of LLMs \cite{lin2023cct5, wang2022no, liu2024delving}. In addition, we thoroughly review all testing results and find that exactly no LLM-generated commit message is identical to the ground truth. Meanwhile, our proposed framework, namely ERICommiter, also achieves substantial improvement over its underlying LLMs. Considering all the factors above, we believe the threat of data leakage is limited. 

\textbf{External Validity:} One potential threat to external validity is that we do not exhaustively evaluate all existing LLMs, but selected four LLMs for evaluation. However, among the LLMs we selected, there are both open-source LLMs (Gemini, LLaMA) and closed-source LLMs (GPT-3.5), and we also chose LLMs with different parameter sizes (LLaMA-7B and LLaMA-13B). Besides, LLMs of diverse families are also included (GPT-3.5, LLaMA, and Gemini). Therefore, our evaluation of LLMs is relatively representative and can mitigate this threat. In the future, we can involve more LLMs for experiments.

Another threat to external validity concerns certain manual errors or inaccuracies during the data cleaning process for MCMDEval+. To address this, we exclude non-target PL samples by regular expressions and harness a deep learning-based model to conduct further filtering semantically to improve the density of high-quality samples, thereby making the follow-up manual selection more concentrated on high-quality samples and improving the selection efficiency and efficacy. In addition, we introduce a pairwise evaluation mechanism to manually select high-quality commit messages, where kappa consistency scores among the evaluators are all over the required criterion, demonstrating a rigorous and convincing procedure. Hence, we believe this threat can be minimal.

\textbf{Construct Validity:} The lack of comprehensive evaluation metrics also poses a threat to validity. In this study, we select METEOR, BLEU, and ROUGE-L as the assessment metrics for CMG approaches and LLMs because they are widely used in the CMG domain. Additionally, aligning with the OSS practice, we also propose four manual evaluation aspects, covering Accuracy, Integrity, Readability, and Applicability to systematically measure different kinds of approaches' CMG abilities. Therefore, the evaluation metrics adopted in this paper are representative and comprehensive.

\section{Conclusion}
This paper serves as the first systematic study investigating the capability of LLMs in generating commit messages based on given code \textit{diff}s. 
Specifically, motivated by our pilot analysis, we first screen and construct a high-quality test set, namely MCMDEval+, through a rigorous three-step cleaning based on two Commit Message Generation (CMG) datasets, i.e., MCMD and CommitChronicle.
%the most widely used Commit Message Generation (CMG) dataset, namely MCMD. 
Afterward, we assess the CMG performance of recent LLMs against state-of-the-art CMG approaches of diverse categories and carry out an in-depth manual analysis from aspects of Accuracy, Integrity, Readability, and Applicability, aligning with the Open-Source Software (OSS) practice. Results demonstrate the superiority of LLMs in the CMG task where GPT-3.5 performs the best. Finally, we further propose an Efficient Retrieval-based In-context learning framework, namely ERICommiter, to improve LLMs' CMG performance. Comprehensive experiments prove the remarkable efficacy and efficiency of ERICommiter, showing its practicality in OSS development and maintenance.

\textbf{Implications for Practitioners:}
The research highlighted the significant benefits of using LLMs for CMG in the OSS practice. By re-evaluating state-of-the-art CMG approaches and LLMs, we presented a more objective and authentic result for practitioners to instruct their applications of CMG approaches in daily development and maintenance.
Besides, since mainstream LLMs offer corresponding APIs, practitioners do not need to invest substantial computational resources to deploy LLMs locally.       Additionally, our proposed ERICommitter substantially enhances LLMs' CMG performance in a training-free manner, boasts a low time overhead and the retrieval component incurs minimal computational resource consumption, making it an effective and efficient LLM-based CMG approach that can be extensively deployed in modern software development practice.

\textbf{Implications for Researchers:}
This work carried out the first systematic empirical study on LLMs' performance in the CMG field. A series of automatic and manual assessments demonstrate their prospects and limitations, shedding light on the research of future alternative approaches. Besides, we constructed a high-quality test set, namely MCMDEval+, cleaned from the most widely used CMG dataset (i.e., MCMD) and the latest CMG dataset (i.e., CommitChronicle), facilitating researchers to evaluate future CMG approaches from a more practical perspective. Finally, our proposed framework, namely ERICommiter, leveraging retrieval and ICL techniques, highly promotes the advancement of CMG approaches in both efficacy and efficiency, proving the potential of ICL in generating commit messages.  

Future research should focus on extending LLMs' capabilities to generate coherent commit messages from multiple code \textit{diff}s, potentially across various programming languages, to broaden their utility in heterogeneous development contexts. Besides, future research can also investigate more advanced prompt designs (e.g., incorporating chain-of-thought techniques) to further enhance the performance of LLMs, thereby ensuring their sustained impact and effectiveness in the ever-evolving landscape of open-source software.

% \section*{Acknowledgments}
% This work is partially supported by the National Natural Science Foundation of China (Grant No. 62192731, 62102233), Natural Science Foundation of Shandong Province (Grant No. ZR2024QF093), Shandong Province Overseas Outstanding Youth Fund (Grant No. 2022HWYQ-043). 

% {\appendix[Proof of the Zonklar Equations]
% Use $\backslash${\tt{appendix}} if you have a single appendix:
% Do not use $\backslash${\tt{section}} anymore after $\backslash${\tt{appendix}}, only $\backslash${\tt{section*}}.
% If you have multiple appendixes use $\backslash${\tt{appendices}} then use $\backslash${\tt{section}} to start each appendix.
% You must declare a $\backslash${\tt{section}} before using any $\backslash${\tt{subsection}} or using $\backslash${\tt{label}} ($\backslash${\tt{appendices}} by itself
%  starts a section numbered zero.)}


% \vspace{-0.27cm}
\bibliographystyle{IEEEtran}
\bibliography{hello}

\end{document}